\documentclass[preprint,12pt,3p]{elsarticle}

\usepackage{amssymb}

\usepackage{lipsum} 	
\usepackage{graphicx}	
\usepackage{float}		

\usepackage{color,soul}
\usepackage{booktabs}
\usepackage[normalem]{ulem}
\useunder{\uline}{\ul}{}

\usepackage{longtable,booktabs}
\usepackage{multirow}
\usepackage[table,xcdraw]{xcolor}
\usepackage{mathtools}
\usepackage{amssymb}
\usepackage{amsmath}
\usepackage{caption}
\usepackage{subcaption}
\usepackage{tikz}

\usepackage{arydshln} 

\makeatletter
\newcommand{\vast}{\bBigg@{4}}
\newcommand{\Vast}{\bBigg@{7}}
\makeatother

\usepackage{framed} 
\usepackage{nomencl} 
\makenomenclature
\usepackage{etoolbox}
\renewcommand\nomgroup[1]{%
  \item[\bfseries
  \ifstrequal{#1}{L}{List of Symbols}{%
  \ifstrequal{#1}{C}{Superscripts}{%
  \ifstrequal{#1}{B}{Subscripts}{}}}%
]}

\begin{document}

\begin{frontmatter}

\title{Secondary breakup of drops at moderate Weber numbers: Effect of Density ratio and Reynolds number}

\author[label1]{Suhas S Jain\fnref{label3}}
\author[label1]{Neha Tyagi}
\author[label1]{R. Surya Prakash}
\author[label1]{R. V. Ravikrishna}
\author[label1]{Gaurav Tomar\corref{cor1}}
\ead{gtom@iisc.ac.in}
\address[label1]{Department of Mechanical Engineering, Indian Institute of Science, India.}

\cortext[cor1]{Corresponding author}
\fntext[label3]{Present address: Department of Mechanical Engineering, Stanford University, USA.}

\begin{abstract}
Breakup of liquid drops occurs in several natural and industrial settings. Fully resolved Volume of Fluid based simulations presented in this study reveal the complete flow physics and droplet dynamics that lead to the breakup of a drop in a particular mode. We have investigated the effects of density ratio and Reynolds number on the dynamics of drop deformation and subsequent breakup. A density ratio-Weber number phase plot is presented that indicates the variation in the deformation of the drop at various density ratios and Weber numbers. We show that the breakup dynamics of the droplets at low density ratios is significantly different to that observed at high density ratios. We also study the temporal characteristics of the droplet deformation and motion.
\end{abstract}


\end{frontmatter}

\section{Introduction} \label{sec:intro}

When a drop is accelerated in a high speed gas flow, it deforms due to the aerodynamic forces and eventually fragments into tiny droplets; this is termed as secondary breakup. This phenomenon has been studied over many decades in the interest of its numerous applications, for example, in rainfall, sprays, combustion and chemical industries. Complete understanding of the breakup phenomenon is essential for an accurate determination of the drop size distribution which dictates the surface to volume ratio and hence the efficiency of drying, chemical reaction and combustion. Further, a better understanding of the breakup also helps in developing accurate closure relations for Lagrangian and Eulerian Multi-Fluid modelling approaches.
    
Over the years, numerous experimental and numerical studies have been performed to study secondary breakup of a drop. Several articles (\cite{Pilch1987,Faeth1995,Guildenbecher2009}) have periodically reviewed the advances in this field. The secondary breakup of a drop can be broadly categorised into four modes of deformation and breakup, primarily based on the aerodynamic Weber number and liquid Ohnesorge number: (a) Vibrational mode, where a drop oscillates at its natural frequency and it may (or may not) undergo breakup \citep{Hsiang1992} and when it breaks it produces fewer daughter drops of the size comparable to that of the parent drop \citep{Pilch1987}, (b) Bag mode, where a drop deforms into a flat disc and then is blown into a thin bag, attached to a toroidal ring, that expands and eventually ruptures, followed by the breakup of the toroidal ring \citep{Chou1998,Jain2015}. Bag fragments into a large number of smaller sized drops and the ring breaks up into smaller number of larger sized drops. With an increasing Weber number, some interesting features appear such as bag with a stamen \citep{Pilch1987} and bag with multiple lobes \citep{Cao2007}. This phenomenon is thought to be due to Rayleigh-Taylor (RT) instability \citep{Theofanous2004,Zhao2010,Jain2015} or a combined RT/aerodynamic drag mechanism \citep{Guildenbecher2009}. (c) Sheet thinning mode for higher Weber number, where the ligaments and small daughter drops break off from the thinning rim of the parent drop until the core of the parent drop reaches a stable state. Earlier, shear stripping (a viscous phenomenon) was assumed to be the mechanism \citep{Ranger1969}, but later \citet{Liu1997} proposed the sheet thinning mechanism, pointing out that it is an inviscid phenomenon. (d) Catastrophic mode, where a drop breaks up into multiple fragments due to unstable surface waves at high speeds \citep{Liu1993}. The transition of this breakup mode, as a function of Weber number, occurs very gradually (see Figure \ref{fig:phaseplot}). Different authors have proposed different transitional values of $We$ (subject to the inaccuracies in the exact calculation of $We$ and also the presence of impurities that alter the properties of the fluid used in the experiments; see Table \ref{tab:transitional_we}) and hence the reliability of the transitional values of $We$ has remained a moot point. Other parameters that influence the breakup mechanism are density ratio and gas Reynolds number (and liquid Ohnesorge number).

Several experimental studies have been performed in the last decades to unravel the physics of secondary breakup of droplets (see \cite{Guildenbecher2009} for an elaborate review). Some of the studies have been performed at low density ratio (in the range $1-10$) while most of the studies have been performed for water-air systems. \cite{Simpkins1972} and \cite{Harper1972} presented experimental and theoretical studies, respectively, of secondary breakup of droplets at high Bond numbers. \cite{Harper1972} showed that Rayleigh-Taylor instability is dominant at higher Bond numbers after a short time algebraic deformation in time.
Experiments of \cite{Simpkins1972} corroborated the theoretical findings. \cite{Patel1981} studied the fragmentation of drops moving at high speed for mercury/water system (density ratio $\sim 10$). They showed that the breakup time even at low density ratios correlates well with the time constant for the growth of unstable Taylor waves in the entire range of Bond numbers. Later, \cite{Theofanous2004} studied droplet breakup at different static pressures over a wide range of gas densities, all in the rarefied range. In this study, they noted that the upstream face of the droplet, upon droplet flattening, becomes immediately susceptible to Rayleigh-Taylor instability. Further, they showed that for low Weber numbers, Bag formation corresponds to one Rayleigh-Taylor wave in the 'disc-shaped' droplet. \cite{Lee2000} performed experiments for a range of gas densities (corresponding to density ratios $100-1000$). Using a pressurized chamber the ambient air pressure was controlled to vary the density ratio from 100 to 1000. They concluded that the Rayleigh number (and also the density ratio) have little effect on the drop breakup mechanisms, although the transition Weber numbers vary a bit. \cite{Gelfand1996} studied liquid(drop)-gas and dense liquid(drop)-light liquid systems and discussed the similarities between the features observed in the two systems. He noted that the value of the first critical Weber number in liquid-liquid systems is higher than the gas-liquid systems. The breakup features, bag and bag-stamen, observed at low Weber numbers (around 15 to 40) for gas-liquid systems, Gelfand noted, do not appear clearly in the liquid-liquid systems.

\begin{table}
\begin{center}
\resizebox{\textwidth}{!}{
\begin{tabular}{cccccccc}
\textbf{Breakup regime}                                                   & 1            & 2          & 3                                                                                 & 4                                                                                 & 5          & 6          & 7           \\[3pt] 
Vibrational                                                      & $We<12$      & $We<10$    & $We<11$                                                                           & $We<13$                                                                           &            &            & $We<12$     \\
Bag                                                              & $12<We<50$   & $10<We<18$ & $11<We<35$                                                                        & $13<We<35$                                                                        & $13<We<18$ &            & $12<We<24$  \\
Bag-stamen                                                       & $50<We<100$  & $18<We<30$ & \multirow{4}{*}{\begin{tabular}[c]{@{}c@{}}$35<We<80$\\ (multimode)\end{tabular}} & \multirow{4}{*}{\begin{tabular}[c]{@{}c@{}}$35<We<80$\\ (multimode)\end{tabular}} &            &            & $24<We<45$  \\
Bag-plume                                                        &              &            &                                                                                   &                                                                                   & $18<We<40$ &            & $45<We<65$  \\
Multibag                                                         &              &            &                                                                                   &                                                                                   &            & $28<We<41$ & $65<We<85$  \\
Plume-shear                                                      &              &            &                                                                                   &                                                                                   & $40<We<80$ &            &             \\
\begin{tabular}[c]{@{}c@{}}Sheet (Shear)\\ thinning\end{tabular} & $100<We<350$ & $We>63$    &                                                                                   & $80<We<800$                                                                       &            &            & $85<We<120$ \\
Catastrophic                                                     & $We>350$     &            &                                                                                   & $We>800$                                                                          &            &            &             \\[6pt]
\multicolumn{8}{l}{ \begin{tabular}[c]{@{}c@{}}
1. \citet{Pilch1987}, 2. \citet{Krzeczkowski1980}, 3. \citet{Hsiang1995}, 4. \citet{Chou1997}, 
5. \citet{Dai2001}, \\ 6. \citet{Cao2007}, 7. \citet{Jain2015}
\end{tabular}}
\\ 
\end{tabular}}
\caption{Transition Weber number ($We$) for different breakup regimes.}
\label{tab:transitional_we}
\end{center}
\end{table}


Most of the numerical studies have been performed with low density ratios and only a few with high density ratio. Efforts in numerical studies have only started to pay off recently and most numerical simulations attempt to study the breakup at low density ratios ($\rho^*<100$), essentially due to numerical convergence issues at high density ratios. Nevertheless, these studies find direct applications in high-pressure environment applications as well as in manufacturing of metal pellets by quenching liquid metal droplets. \citet{Zaleski1995} performed one of the earliest numerical studies on the secondary breakup of drops in 2D. They observed a backward-bag at low Weber number ($We$) for $\rho^*=10$ and reported that their results contradict the general experimental observation (which was mostly done for higher $\rho^*$ values), where a forward-bag is seen at this $We$. They suggested that this mismatch is a result of the discrepancy in their initial conditions.  \citet{Han2001} extensively studied the breakup of drops for two $\rho^*$ values, 1.15 and 10. For $\rho^*=10$, they observed a forward-bag at low $We$ and backward-bag at higher $We$, and for $\rho^*=1.15$ they observed backward-bag for all moderate $We$. They concluded that the formation of forward-bag is due to the detachment of the wake downstream of the drop and the formation of backward-bag is due to the entrapment of the drop in the vortex ring. On decreasing $Re$, they also observed that a higher $We$ is required to obtain the same mode of breakup. \citet{Aalburg2003} reported that the secondary atomization is essentially independent for $\rho^*>32$ and that there is no effect of $Re$ on $We_{crit}$ beyond $Re>100$. \citet{Kekesi2014} studied the breakup of drops for $\rho^*=$20, 40 and 80 and reported to have observed new breakup modes such as Bag, Shear, Jellyfish shear, thick rim shear, thick rim bag, rim shear and mixed. The new breakup modes were due to the influence of the viscous effects in their simulations (some of these cases are at $Oh\gg0.1$ and $Re_g < 100$). \citet{Yang2016} also studied the effect of $\rho^*$ on the breakup but at a very high $We=225$ value in the regime of catastrophic breakup for $\rho^*$=10, 25, 32, 60. On decreasing $\rho^*$, they observed a lower deformation rate but the range of $\rho^*$ values chosen was probably too low at such high $We$ to see any discernible effect of changing $\rho^*$ on the breakup. Formation of spherical cap and ligaments and the fragmentation of ligaments further into multiple drops were the common features they observed in their study. 
Recently, 3D simulations were performed for water and air at atmospheric conditions ($\rho^*\sim$1000) by \citet{Xiao2014,Xiao2016}; but their main focus was to validate their LES code. We, in our previous work \citep{Jain2015}, have extensively studied the breakup and its characteristics for $\rho^*=1000$ using fully resolved 3D simulations. 



For the systems with low density ratios ($<100$) and at moderate Weber numbers (20-80), backward-bag (opening of the bag facing the downstream direction followed by sheet thinning) has been seen  as the predominant breakup mode in the numerical simulations (see \citet{Kekesi2014,Khosla2006}). 
In the present work, we numerically study the effect of a wide range of density ratios on the drop breakup mechanisms at different aerodynamic Weber numbers. We focus on the effect of flow features in the surrounding medium on drop deformation and its subsequent fragmentation. 

The paper is organized as follows. Section \ref{sec:problem} describes the equations solved, the computational domain and the grid independence study. Results and discussion are presented in Section \ref{sec:result}.
The summary of the study and important conclusions are discussed in Section \ref{sec:conclude}.

\section{Problem description and Formulation} \label{sec:problem}

\begin{figure}
\centering
\includegraphics[width=0.45\textwidth]{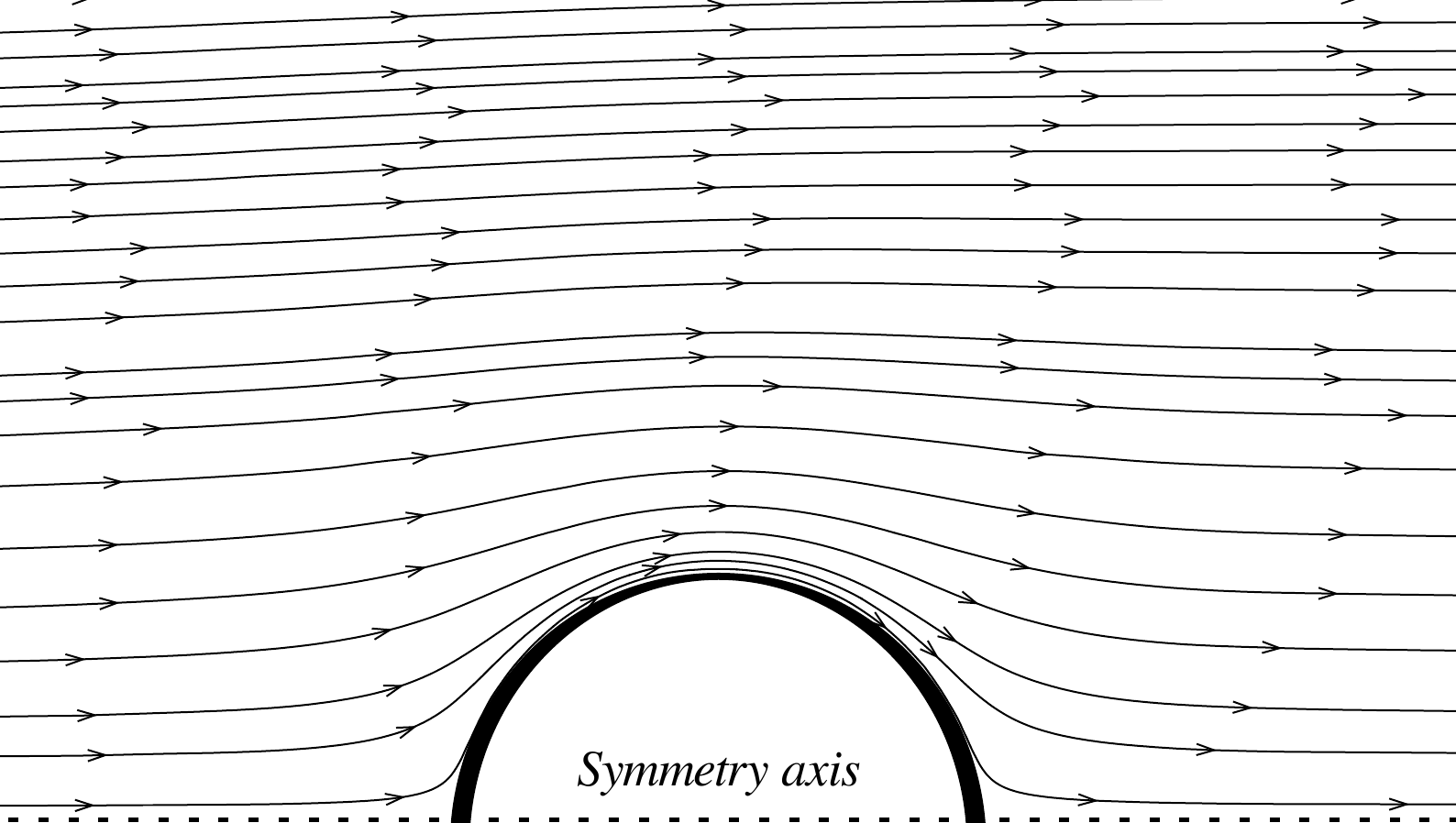}
\caption{Schematic of the simulation setup (not drawn to scale). Flow direction represents the direction of the gas flow. Drop is shown at the initial time ($t=0$).}
\label{fig:domain}
\end{figure}

Figure \ref{fig:domain} shows the schematic of the computational domain for the axisymmetric simulations performed in this study with the dashed line marking the axis of symmetry.
The domain is $10d_0$ along the radial direction and $20d_0$ along the axial direction, where $d_0$ is the diameter of the drop.
Liquid and gas densities are $\rho_l$ and $\rho_g$, respectively, and the ratio $\rho^* = \rho_l/\rho_g$ is varied from $10$ to $1000$ by keeping the gas density as unity and varying the liquid density. Viscosity of the liquid and the gas are given by $\mu_l$ and $\mu_g$, respectively. Surface tension coefficient at the liquid-gas interface is given by $\sigma$.
 
For the simulations, gas inlet is at the left and is prescribed a uniform velocity of $U_{g}$, and outlet flow boundary conditions are imposed at the right end of the computational domain. Slip boundary conditions are applied at the other (side) walls of the domain to minimize the confinement effects and axisymmetric boundary conditions are imposed at the axis of symmetry marked by the dashed line in Figure \ref{fig:domain}.
The drop is accelerated by the high-speed gas flow and its breakup is governed by the following five non-dimensional numbers: Aerodynamic Weber number $We=\rho_g U_{g}^2 d_0/\sigma$ (ratio of the gas inertial forces to the surface tension forces), liquid Ohnesorge number $Oh=\mu_l/\sqrt{\rho d_0 \sigma}$ (ratio of drop viscous forces to the surface tension and the drop inertial forces), gas Reynolds number $Re=\rho_g U_{g} d_0/\mu_g$ (ratio of gas inertial forces to the gas viscous forces), viscosity ratio $M=\mu_l/\mu_g$ (ratio of drop viscosity to the gas viscosity) and the density ratio $\rho^*=\rho_l/\rho_g$ (ratio of drop density to the gas density).



A one-fluid formulation is used for the numerical simulations \citep{Mirjalili2017}. The governing equations for the coupled liquid and gas flow simulated in this study are described in the following.
Considering both the drop fluid and the surrounding gas to be incompressible, the corresponding continuity equation is given by,
\\
\begin{equation}
\nabla \cdot \mathbf{u}=0,
\label{equ:d_continuity}
\end{equation}
\\
where $\mathbf{u}$ is the divergence free velocity field. The governing equations for the momentum conservation are given by the Navier$-$Stokes equations (Eqn. \ref{equ:NS}) modified to implicitly account for the surface tension forces and the interfacial boundary conditions of continuity of velocity, and normal and tangential stress balance:
\\
\begin{equation}
\rho(C)(\frac{\partial\mathbf{u}}{\partial t}+\nabla \cdot \mathbf{uu})=-\nabla p + \nabla \cdot (2\mu(C) \mathbf{D}) + \sigma \kappa \mathbf{n} \delta_{s}.
\label{equ:NS}
\end{equation}
\\
Here, $C$ is the volume fraction of liquid that takes a value of zero in the gas phase and one in the liquid phase. The density and viscosity for the one-fluid formulation are expressed as, $\rho=\rho_{l}C+\rho_{g}(1-C)$ and  $\mu=\mu_{l}C+\mu_{g}(1-C)$, respectively.
The deformation rate tensor is given by $D=(\nabla \textbf{u}+(\nabla \textbf{u})^{T})/2$. The last term in the equation ($\sigma \kappa \textbf{n} \delta_s$) accounts for the surface tension force ($\sigma \kappa$, where $ \kappa $ is the local interface curvature) acting on the interface, expressed as a volumetric force using the surface Dirac delta function ($\delta_{s}$) and modeled using the continuum surface force approach \citep{Brackbill1992}. The direction of this force is along the local normal $(\textbf{n})$ at the interface. The evolution equation for the interface is given as an advection equation in terms of the volume fraction, $C$  (obtained by applying kinematic boundary condition at the interface),
\\
\begin{equation}
\frac{\partial C}{\partial t} + \textbf{u}.\nabla C = 0.
\label{equ:advection}
\end{equation}
\\

We use a cell-based Octree grid adaptive mesh refinement (AMR) geometric volume of fluid (VOF) algorithm in Gerris \citep[see][]{Popinet2003,Popinet2009,Tomar2010} to solve the above set of equations. Gerris uses a second-order accurate staggered time discretization for velocity, volume-fraction and pressure fields. Balanced-force algorithm by \cite{Francois2006} is used to accurately calculate the surface tension forces and minimize spurious currents. The discretization of the equations (Eqn. \ref{equ:d_continuity}-\ref{equ:d_NS}) are described in detail in \cite{Popinet2003} and will be discussed here only briefly. 

Discretized Navier$-$Stokes equations are solved implicitly using a projection method. First, an auxiliary velocity field is obtained using the following discretization (\cite{Popinet2003}):
\begin{equation}
    \rho_{n+\frac{1}{2}}\Big(\frac{\textbf{u}_* - \textbf{u}_n}{\nabla t} + \textbf{u}_{n+\frac{1}{2}}.\nabla \textbf{u}_{n+\frac{1}{2}}\Big)=\nabla \cdot(\mu_{n+\frac{1}{2}}(\textbf{D}_n + \textbf{D}_*)) + (\sigma \kappa \delta_s \textbf{n})_{n+\frac{1}{2}}.
    \label{equ:d_NS}
\end{equation}

Void fraction is updated using the following equation with the fluxes computed geometrically (\cite{Popinet2009}):
\begin{equation}
    \frac{C_{n+\frac{1}{2}}-C_{n-\frac{1}{2}}}{\Delta t} + \nabla.(C_n \textbf{u}_n)=0.
    \label{equ:d_advect}
\end{equation}

The pressure Poisson equation:
\begin{equation}
\nabla \cdot \left(\frac{\nabla p}{\rho(C)}\right) = \frac{\nabla \cdot \textbf{u}_*}{\Delta t},
    \label{equ:d_project}
\end{equation}
 is solved using a geometric multigrid method and the auxiliary velocity field is updated as following to obtain a divergence free velocity field ( $\nabla \cdot \textbf{u}_{n+1}=0$),
\begin{equation}
    \textbf{u}_{n+1}=\textbf{u}_* - \frac{\Delta t}{\rho_{n+\frac{1}{2}}}\nabla p_{n+\frac{1}{2}}.
    \label{equ:d_project}
\end{equation}



\begin{figure}
\centering
\includegraphics[width=0.45\textwidth]{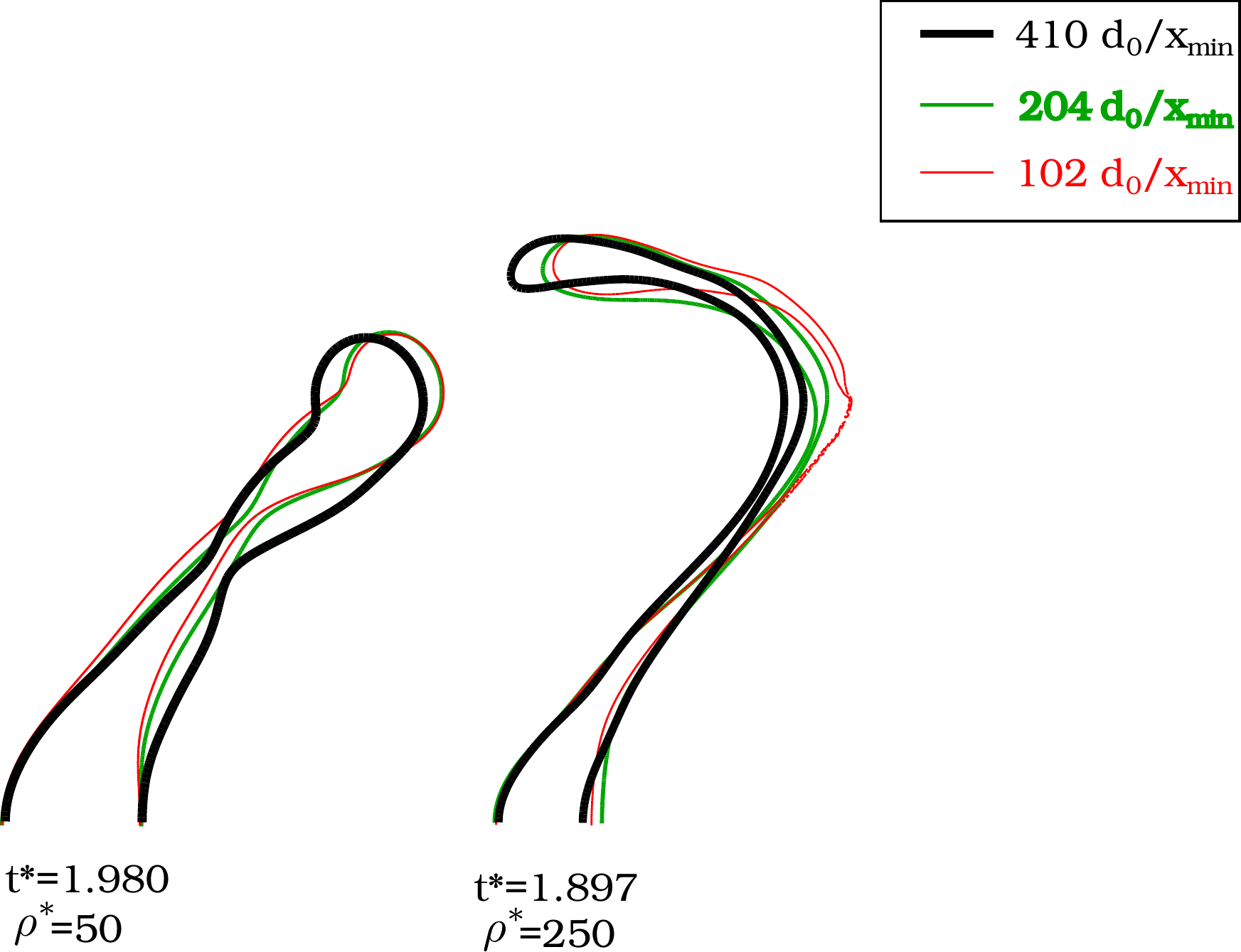}
\centering
\caption{Drop shapes for $\rho^*=50$ and $250$ compared with profiles, obtained using grid resolutions $102, 204$ and $410$ $d_0/\Delta x_{min}$, at the same time instant.}
\label{fig:grid}
\end{figure}

Adaptive mesh refinement (AMR) is performed using a cost function based on the local vorticity in the field and the gradient of the void-fraction field, thus using a very fine refinement in the regions of high velocity gradient and at the interface. We use 410 cells per diameter (${d_0}/{\Delta x_{min}}$) of the initial spherical drop for the refinement of the interface and three different grid resolutions for the refinement of surrounding gas flow - 102, 204 and 410 ${d_0}/{\Delta x_{min}}$ for our 2D axisymmetric simulations. This resolution is more than that employed in any of the previous studies in the literature. For example, \cite{Han2001} used around 100 and \citep{Kekesi2014} used 32 number of grid points per droplet diameter. Figure \ref{fig:grid} shows the drop shapes at different grid refinements and density ratios. 

Drop shapes for grid resolutions 102, 204 and 410 ${d_0}/{\Delta x_{min}}$ are identical (in a physically meaningful way), but we use the most fine mesh refinement of 410 ${d_0}/{\Delta x_{min}}$ for all the axi-symmetric simulations presented in this study. We note that the  liquid Reynolds number for the high density ratio cases is above 10000. However, for $Re_l < 10,000$, we note that even a coarser mesh of 102 ${d_0}/{\Delta x_{min}}$ is sufficient. Thus, in order to maintain high-fidelity of the simulations, we performed all simulations choosing $Re_l$ sufficiently smaller than $10,000$. We also perform a few 3D simulations using a mesh refinement of 102 ${d_0}/{\Delta x_{min}}$ to show the validity of our axisymmetric assumption.




In order to perform high density ratio simulations, as discussed in \citep{Jain2015}, since a sharp interface (single grid transtion region) results in an inexplicable spike in the kinetic energy, we use a thin transition region for better convergence. We use a transition region of two cells for $\rho^* \geq 500$  and one cell for $\rho^* \geq 100$ on either sides of the interface for smoothing the jump in the physical properties. To test the efficacy of the numerical algorithm in capturing high density ratios, mainly regarding the use of a thin smoothing width at the interface, we had presented validation test cases in \citep{Jain2015}, which show good agreement with the corresponding analytical results for a density ratio of 1000. In the following section, we discuss the results for a wide range of density ratios on the dynamics of deformation and breakup of drops.

\section{Results and discussion} \label{sec:result}

In order to investigate the effects of density ratio, $\rho^*$, on the secondary breakup of a drop, we perform a large set of well resolved simulations with different values of $\rho^*$, Reynolds number $Re$ and aerodynamic Weber number, $We$. Table \ref{tab:cases} lists the parameter range covered in this study. For each value of $\rho^*$ listed in the table, we vary the Weber number to study the effects of density ratio on the drop breakup regimes, namely, Bag breakup, Bag with Stamen, multi-bag breakup and sheet-thinning breakup which are observed experimentally \citep{Guildenbecher2009} for the range of $We$ chosen here for the simulations. The liquid Ohnesorge number for all the simulations is $\le 0.1$ and therefore the critical Weber number based on the previous studies \citep{Hsiang1992, Krzeczkowski1980} is not expected to vary significantly.


\begin{table}
\begin{center}
\begin{tabular}{c|c|c|c}
$\rho^*$ & $We$  & $Re$ & $M$ \\[3pt] 
\hline \begin{tabular}[c]{c}10, 50, 100, 150,\\ 200, 250, 500, 1000\end{tabular} & \begin{tabular}[c]{c}20, 40, 60,\\ 80, 100, 120\end{tabular} & 4000   & 100             \\ \hdashline 
1000 & 20 & 141, 500, 1414 & 100             \\ \hdashline
1000 & 20 & 1414, 4000, 6000 & 1000         \\ \hdashline
10 & \begin{tabular}[c]{c}20, 40, 60,\\ 80, 100, 120\end{tabular} 
& 20000 & 100 \\ 
\end{tabular}
\caption{Parameters for the different simulations presented in this study.}
\label{tab:cases}
\end{center}
\end{table}


As discussed earlier in Sec.\ref{sec:intro}, there are conflicting views on the effect of density ratio on the breakup mechanism for a given $We$. Where \citet{Aalburg2003} reported that there is little effect of density ratio (for $\rho^* > 32$) on the breakup mechanism, \citet{Jing2010} presented simulations for density ratios 10, 100 and 1000 and showed that a change in the density ratio alters the critical Weber numbers for the different breakup regimes. Several simulations have been reported for low density ratios where for even a reasonably lower $We$, backward bag was observed (see, for example, \citep{Han2001},\citep{Khosla2006}). In this section, we present simulations for a wide range of density ratios and discuss the physical mechanisms that alter the breakup modes.

\begin{figure}
    \centering
    \begin{subfigure}[t]{0.5\textwidth}
        \centering
        \includegraphics[height=1.2in]{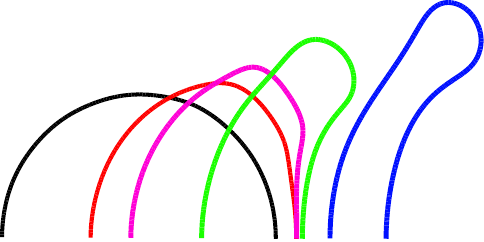}
        \caption{$\rho^*=10$}
        \label{fig:rm_dr10}
    \end{subfigure}%
    ~ 
    \begin{subfigure}[t]{0.5\textwidth}
        \centering
        \includegraphics[height=1.2in]{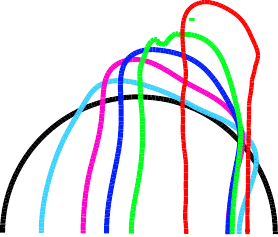}
        \caption{$\rho^*=1000$}
        \label{fig:rm_dr1000}
    \end{subfigure}
    \caption{Time evolution of the drop movement for $\rho^*=10$ and $\rho^*=1000$ at $We=20$.}
    \label{fig:rear_movement}
\end{figure}

In what follows, we discuss the effect of density ratio on the deformation and motion of the droplet. The time is non-dimensionalized with the characteristic time scale, $t^* = t/t_c$, where,
\begin{equation}
t_c = \frac{d_0\sqrt{\rho_l/\rho_g}}{U_{g}}.
\label{def:tc}
\end{equation}
The characteristic time scale is defined by using the diameter of the drop as the characteristic length scale and the velocity scale in the drop obtained by comparing the dynamic pressures in the gas and the drop.
Figure \ref{fig:rear_movement} shows the time evolution of the drop shape and the displacement for $\rho^*=10$ and $\rho^*=1000$ at $We=20$. The flow of the gas is from left to right. Centroid of the drop for $\rho^*=10$ moves a distance of $0.79d_0$, for $\rho^*=200$ drop moves a distance of $0.48d_0$ (not shown in the figure) and for $\rho^*=1000$ the drop moves a distance of $0.34d_0$ in $t^*=1$.  The leeward side of the drop for $\rho^*=10$ also moves downstream with time, whereas the leeward side of the drop for $\rho^*=1000$ remains virtually stationary until $t^*\sim 1$, though the centroid is moving in the streamwise direction in both the cases. The significant difference in the motion of the centroid is primarily due to the differences in the velocity of the drop and the rate of momentum transmitted to the leeward side of the drop, which depends on the kinematic viscosity, $\nu$. The value of $\nu$ for $\rho^*=10$ is 100 times the $\nu$ for $\rho^*=1000$. We can also observe the formation and motion of the capillary waves emanating from the rim of the drops in both the cases (more evidently for the drop at $\rho^*$=1000). Capillary time-scale based on inertia, also called as the Rayleigh time-scale \citep{Rayleigh1879}, is given by $t_R\sim \sqrt{\rho_l d^3/\sigma}$. This is around 3 times the characteristic time scale $t_c$ of the drop in both the cases, since both $t_c$ and $t_R$ are proportional to $\sqrt{\rho_l}$. Capillary time-scale based on viscous forces is given by $t_M\sim \mu d/\sigma$. This is around 0.16 times $t_c$ for the drop with $\rho^*=10$ and around 0.016 times $t_c$ for the drop with $\rho^*=1000$, thus suggesting a lower resistance to the waves by the viscous forces in the case of high density ratio fluid relative to the low density case since $\nu$ is smaller. The stretching time-scale (for the rim) is obtained by the scaling $t_s\sim d/u_{rim}$. This is around 0.76 times the $t_c$ for the drop with $\rho^*=10$ and 1.12 times the $t_c$ for the drop with $\rho^*=1000$. Comparing these time scales, we note that the capillary reorganization occurs at a rapid rate in the high-density ratio case in comparison to the low density ratio cases. Further, the time scale for the stretching of the rim is significantly slower than the capillary wave time-scale for high density ratio cases. Therefore, for higher density ratio cases, a flat disc shape of the droplet is observed, whereas, for lower density ratios, drop progressively deforms into a backward bag without achieving a proper flat disc shape. Note that our definition of backward bag is the one where the rim of the bag is stretched in the direction of the flow relative to the bag. This is different from the one proposed by \cite{Han1999} but is consistent with the one used in \cite{Jain2015}.

\begin{figure}
\centering
\includegraphics[width=0.8\textwidth]{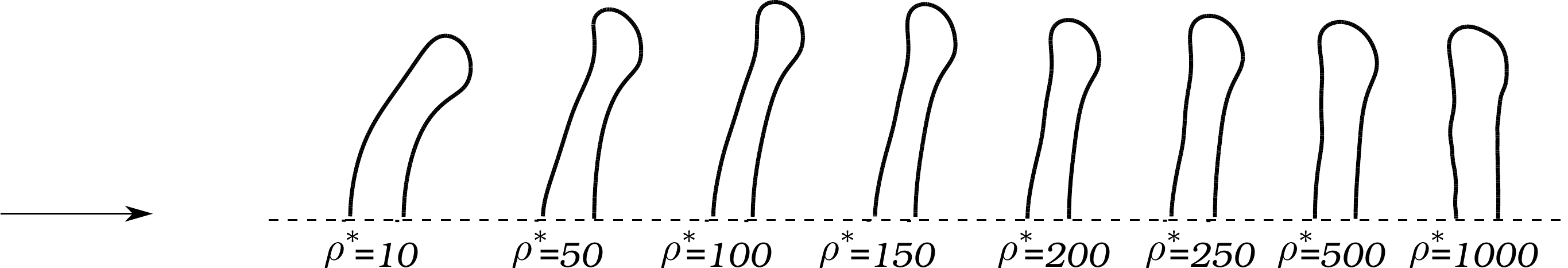}
\centering
\caption{Drop shapes at $t^*=1$ for $\rho^*=$10, 50, 100, 150, 200, 250, 500 and 1000 (left to right) and $We=20$. Arrow denotes the direction of gas flow.}
\label{fig:char_time}
\end{figure}

Figure \ref{fig:char_time} shows the drop shapes at $t^*=1$ for $We = 20$ and for different values of $\rho^*$. 
The flow of the gas is from left to right. We note that the drops at high density ratios deform into a flat disc at around $t^* = 1$ from an initially spherical shape. At low density ratios ($\rho^* = 10$), the formation of disc is not observed at all (see Fig.\ref{fig:rm_dr10}). For intermediate density ratios, the extent of bending of the disc progressively decreases with increase in $\rho^*$ and a near flat disc is obtained for $\rho^* = 1000$. We would like to note here that although $t^* = 1$ is the same for all the profiles shown in Fig.\ref{fig:char_time}, the dimensional time varies as $\frac{1}{\sqrt{\rho^*}}$.
This implies that due to higher velocities, $U_l \sim \sqrt{\frac{1}{\rho^*}} U_g$, obtained during momentum transfer at lower density ratios, the physical time corresponding to $t^* = 1$ is smaller in comparison to higher density ratios. Drop shape for $\rho^* = 10$ is of nearly uniform thickness whereas for intermediate density ratios ($50 < \rho^* < 150$), the drops become thinner near the rim. The thickness at the center of the drop initially decreases and beyond $\rho^* = 150$ it increases slightly and saturates whereas, the diameter of the disc initially increases and beyond $\rho^* = 150$ it undergoes a sudden decrease and saturates (see Fig.\ref{fig:char_time}). The variation in the curvature of the drop with increasing $\rho^*$ suggests a decreasing tendency of forming a backward bag. 







\begin{figure}
\centering
\includegraphics[width=\textwidth]{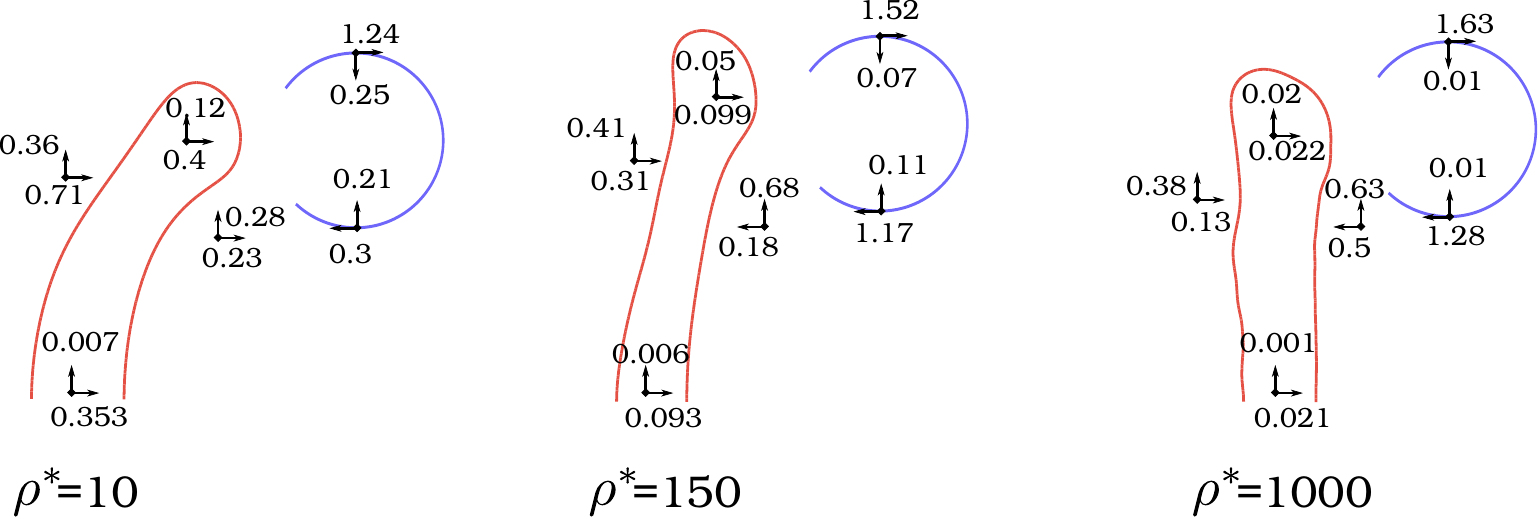}
\centering
\caption{Velocity components, non-dimensionalized using $U_g$, at certain points in and around the drop. Red line represents the drop profile and the blue line represents the vortex behind the droplet. Arrows represent the velocities and are drawn according to the direction but not to scale. Magnitude of the velocity components is mentioned next to the arrows.}
\label{fig:interflow}
\end{figure}

Figure \ref{fig:interflow} shows the non-dimensional (with respect to $U_g$) velocities of the drop and gas at various locations for the drops for $\rho^*=$10, 150 and 1000 at $t^* = 1$. As expected (based on the scaling relation $U_l \sim \sqrt{\frac{1}{\rho^*}} U_g$), the axial and radial components of the velocity at the center ($u_{center}$) and at the rim ($u_{rim}$) of the drop decrease with an increase in $\rho^*$. A vortex is formed behind the drop due to the flow separation as shown in the Figure \ref{fig:interflow}. We can see that the velocities of the vortex (strength of the vortex) in the gas flow is increasing with an increase in $\rho^*$ due to the higher relative velocity of the drop ($u_l-U_g$) for higher $\rho^*$ values. 
Interestingly, the axial component of the velocity at the rim of the drop is significantly higher than that at the center of the drop at lower $\rho^*$. But this difference in velocity at the center and rim $u_{center}-u_{rim}$ is decreasing with increase in $\rho^*$, and for $\rho^*=1000$, we can see that the velocities are almost the same at the rim and that at the center. Percentage of the difference in the velocity at the rim and that at the center (relative to the velocity at the center of the drop), $(u_{rim}-u_{center})/u_{center}$, is $13.3\%$ for $\rho^*=10$, $6.5\%$ for $\rho^*=150$ and $\sim 1.2\%$ for $\rho^*=1000$. Hence, there is more stretching of the rim in the direction of the flow for lower $\rho^*$ values. This explains the bending of the drop for lower $\rho^*$ values and the formation of flat disc for higher $\rho^*$ values. 

\begin{figure}
    \centering
    \begin{subfigure}[t]{\textwidth}
        \centering
        \includegraphics[width=\textwidth]{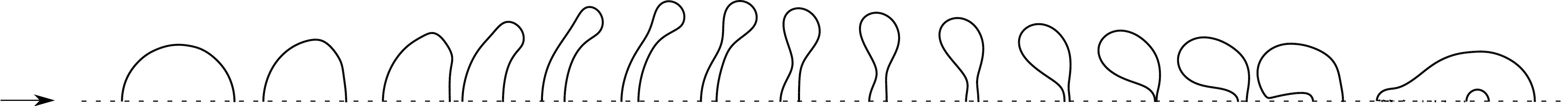}
        \caption{$\rho^*=10, We=20$}
        \label{fig:ev_dr10we20}
    \end{subfigure}
    ~ 
    \begin{subfigure}[t]{\textwidth}
        \centering
        \includegraphics[width=\textwidth]{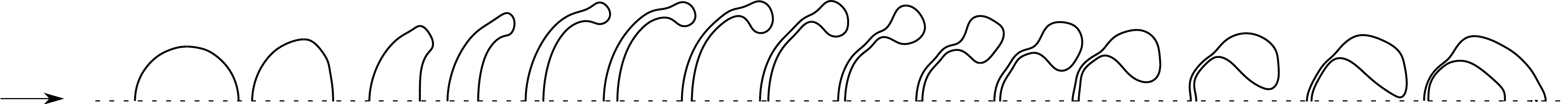}
        \caption{$\rho^*=10, We=40$}
        \label{fig:ev_dr50we20}
    \end{subfigure}
    ~
    \begin{subfigure}[t]{\textwidth}
        \centering
        \includegraphics[width=\textwidth]{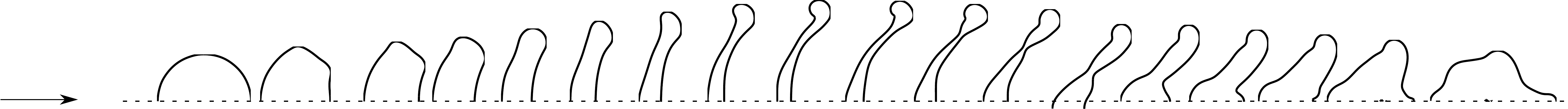}
        \caption{$\rho^*=50, We=20$}
        \label{fig:ev_dr10we40}
    \end{subfigure}
    \caption{Evolution of the drop shape in time for $\rho^*=10$ at $We$=20, 40 and $\rho^*=50$ at $We=20$. Arrows show the direction of gas flow and the dotted lines mark the axis of symmetry. Note that the distance between consecutive droplet profiles plotted here does not represent the actual displacement of the drop.}
    \label{fig:nobreakup}
\end{figure}

Figure \ref{fig:nobreakup} shows the time evolution of the drop shapes for the cases where the drops do not breakup. Drop with $\rho^*=10$ at $We=20$ deforms into a concave-disc facing downstream and then bends in the opposite direction and finally collapses without breakup, encapsulating a bubble within it. For $We=40$ and $\rho^*=10$, it deforms into a backward-bag and again collapses onto itself before it could break. For $\rho^*=50$ at $We=20$, the drop deforms into a concave-disc facing downstream and then into the shape of a canopy-top. Subsequently, with further deformation of the drop, the rim tends to pinch-off from the core drop, but before it could pinch-off, the drop relaxes back collapsing onto itself without breakup. This also shows the highly complicated unsteady behavior of the evolution of drop shapes. To understand this behaviour of no-breakup, we calculate the instantaneous Weber number (based on the velocity of the gas relative to the drop velocity) at the onset of breakup using $We_{inst}=\rho_g (U_g-u_{drop})^2 d_0/\sigma$. Estimating the centroid velocity of the drop $u_{drop}$ from the simulations at $We=20$, we find that the $We_{inst}=3.69$ for $\rho^*$=10, $We_{inst}=8.91$ for $\rho^*$=50 and $We_{inst}=11.1$ for $\rho^*$=100, and $We_{inst}$ increases further with increase in $\rho^*$ value. Clearly for $\rho^*=10$ and $\rho^*=50$, $We_{inst}$ is below the $We_{crit}\sim10-12$, implying that the drop would not breakup. Similarly, at $We=40$, $We_{inst}=8.92$ for $\rho^*$=10 and $We_{inst}=18.31$ for $\rho^*$=50. Here again for $\rho^*$=10, $We_{inst}$ is below the $We_{crit}$ whereas, for $\rho^*$=50, $We_{inst} > We_{crit}$ for an initial aerodynamic Weber number $We = 40$, and thus we observe breakup of the drop. These predictions based on the criterion $We_{inst} > We_{crit}$ for the breakup of drop are in good agreement with our numerical results (as also shown in Figure \ref{fig:nobreakup}).  Thus, we can conclude that the breakup of a drop not only depends on the initial $We$ value but also on the initial dynamics of the drop. More importantly, for low density ratio, for the same momentum transfer the relative velocity decreases much faster in comparison to the rates of deformation of the drop, thus, the instantaneous $We$ decreases sharply and vibrational modes, without breakup, are observed.

\begin{table}
	\begin{center}
	\resizebox{\textwidth}{!}{%
		\begin{tabular}{cc||cccccc}
			\multicolumn{2}{c}{} & \multicolumn{6}{c}{\textbf{$We$}} \\[3pt] 
			\multicolumn{2}{c||}{\multirow{-2}{*}{}} & \textbf{20} &  \textbf{40} & \textbf{60} & \textbf{80} & \textbf{100} & \textbf{120}   \\ \cmidrule{2-8} 
			& \textbf{10} & \raisebox{-0.2in}{\includegraphics[scale=0.3]{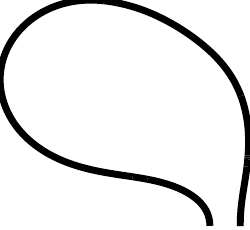}}
			& \raisebox{-0.2in}{\includegraphics[scale=0.3]{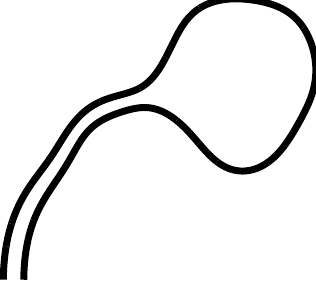}}
			& \raisebox{-0.2in}{\includegraphics[scale=0.3]{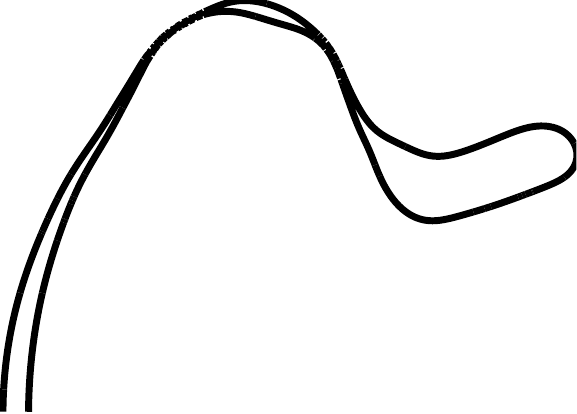}}
			& \raisebox{-0.2in}{\includegraphics[scale=0.3]{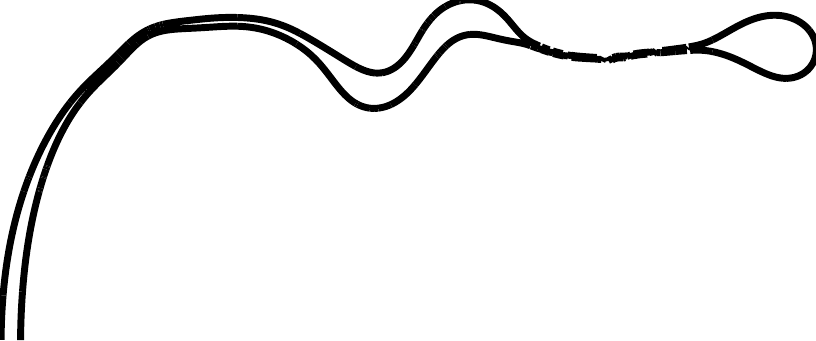}} 
			& \raisebox{-0.2in}{
			 	 \includegraphics[scale=0.3]{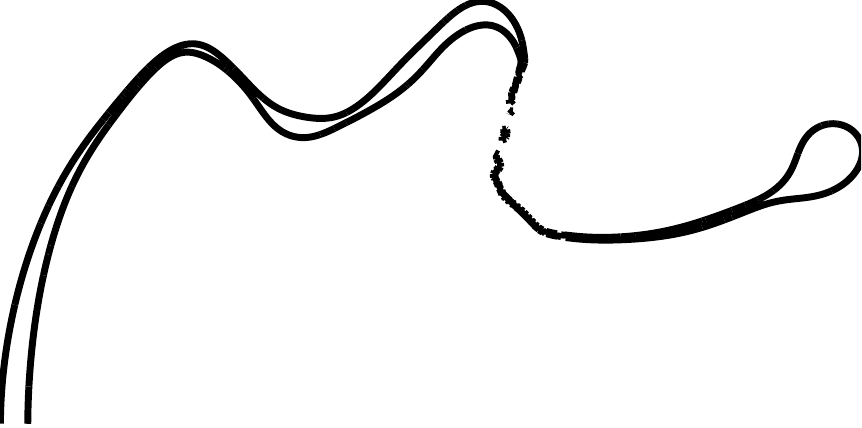}
			  } 
			& \raisebox{-0.2in}{
			 	 \includegraphics[scale=0.3]{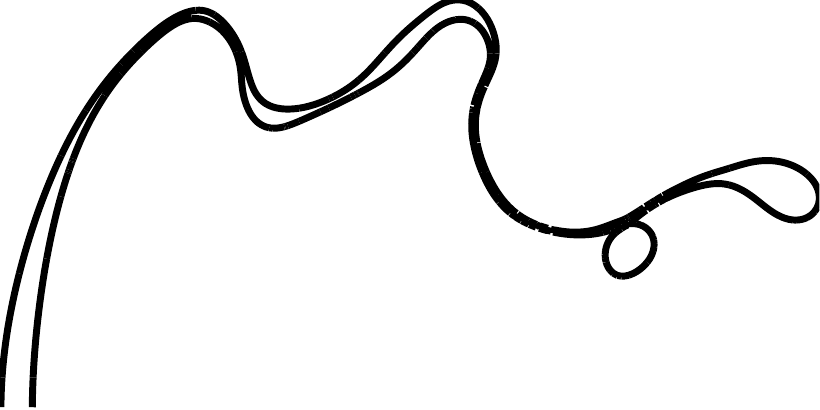}
			  }   \\ \cdashline{3-8} \\[-0.5em]

			& \textbf{50} & \raisebox{-0.2in}{
			 	\includegraphics[scale=0.3]{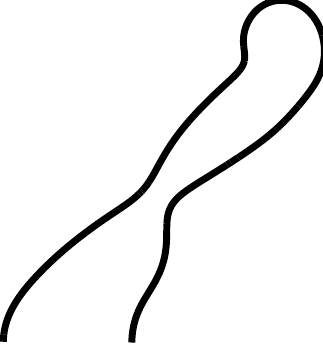}
			 	} 
			& \raisebox{-0.2in}{
			 	 \includegraphics[scale=0.3]{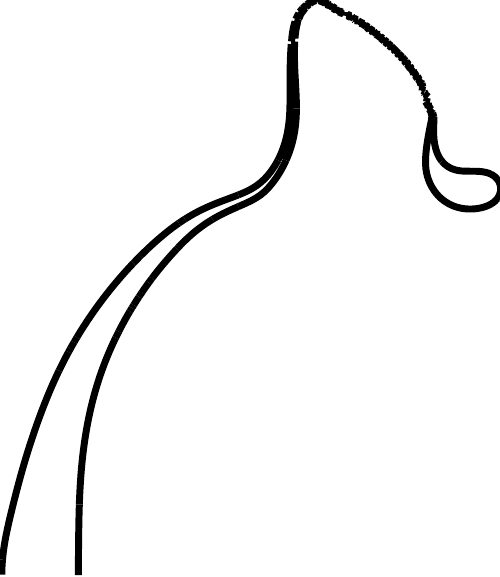}
			  } 
			& \raisebox{-0.2in}{
			 	  	\includegraphics[scale=0.3]{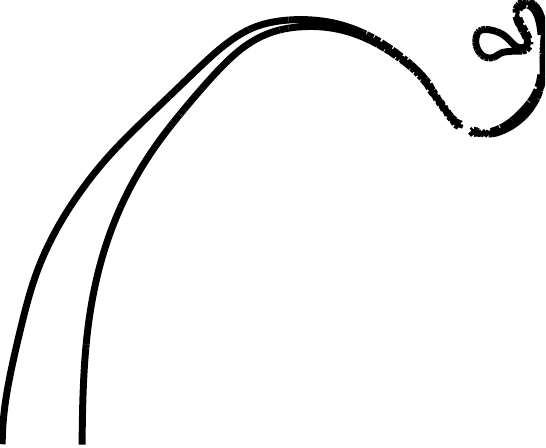}
			 	 } 
			& \raisebox{-0.2in}{
			 	 \includegraphics[scale=0.3]{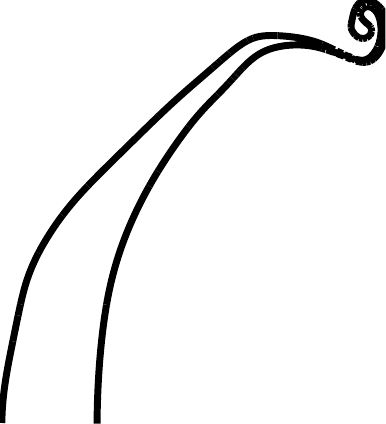}
			  } 
			& \raisebox{-0.2in}{
			 	 \includegraphics[scale=0.3]{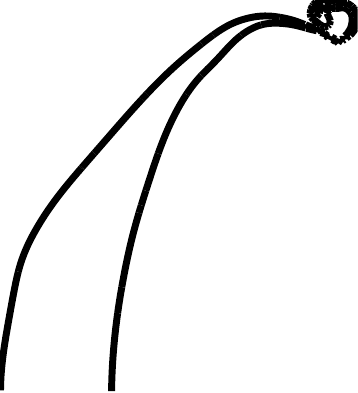}
			  } 
			& \raisebox{-0.2in}{
			 	 \includegraphics[scale=0.3]{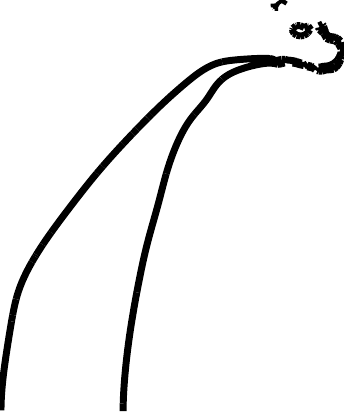}
			  }   \\ \cdashline{3-8} \\[-0.5em]

			& \textbf{100} & \raisebox{-0.2in}{
			 	\includegraphics[scale=0.3]{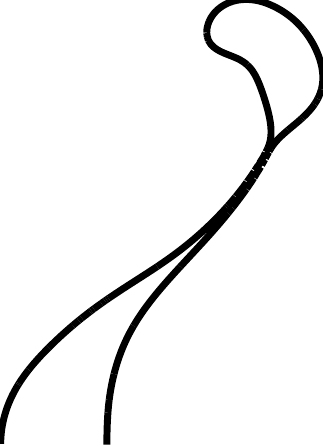}
			 	} 
			& \raisebox{-0.2in}{
			 	 \includegraphics[scale=0.3]{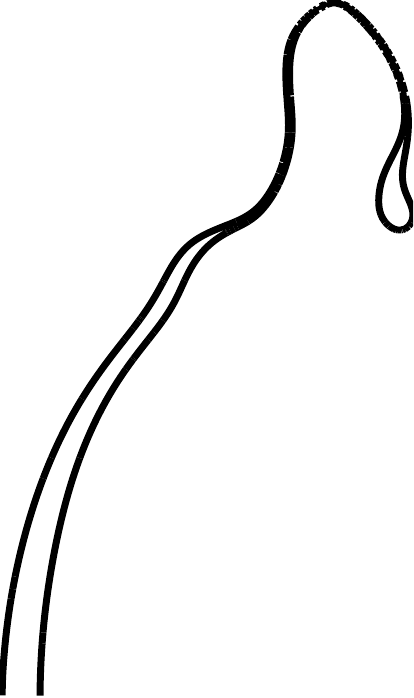}
			  } 
			& \raisebox{-0.2in}{
			 	  	\includegraphics[scale=0.3]{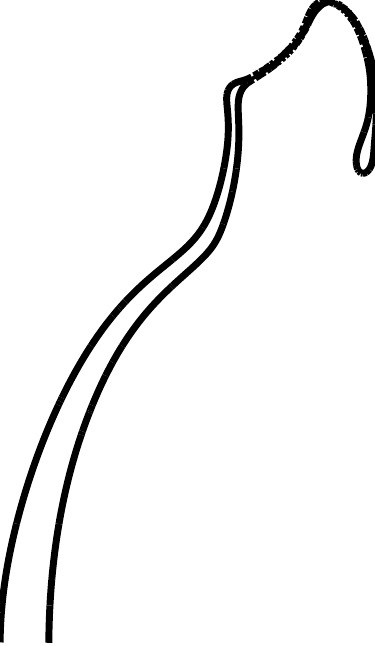}
			 	 } 
			& \raisebox{-0.2in}{
			 	 \includegraphics[scale=0.3]{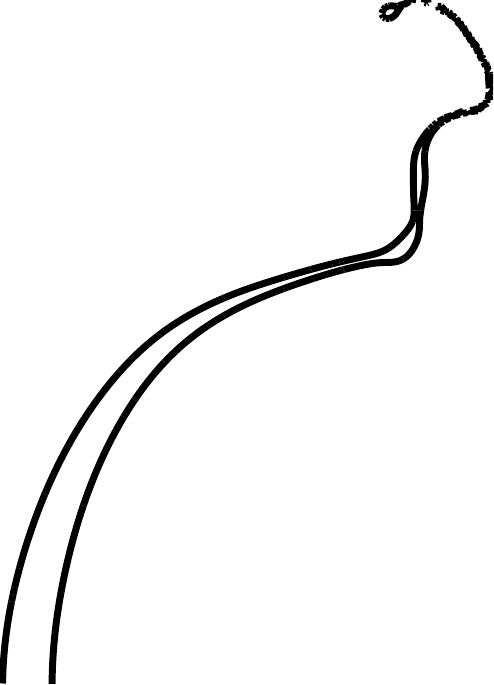}
			  } 
			& \raisebox{-0.2in}{
			 	 \includegraphics[scale=0.3]{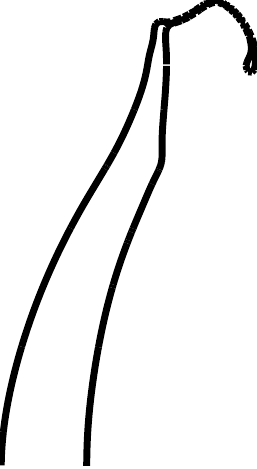}
			  } 
			& \raisebox{-0.2in}{
			 	 \includegraphics[scale=0.3]{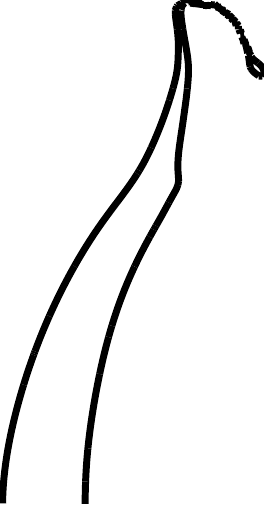}
			  }   \\ \cdashline{3-8} \\[-0.5em]

			& \textbf{150} & \raisebox{-0.2in}{
			 	\includegraphics[scale=0.3]{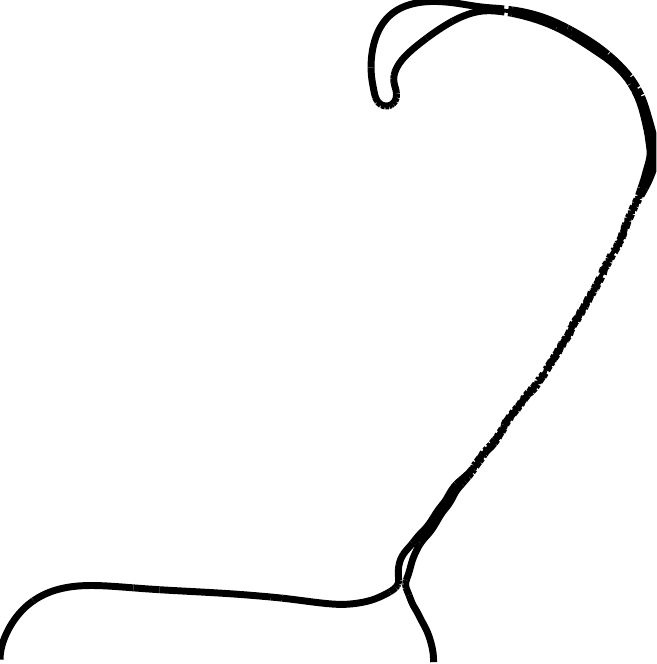}
			 	} 
			& \raisebox{-0.2in}{
			 	 \includegraphics[scale=0.3]{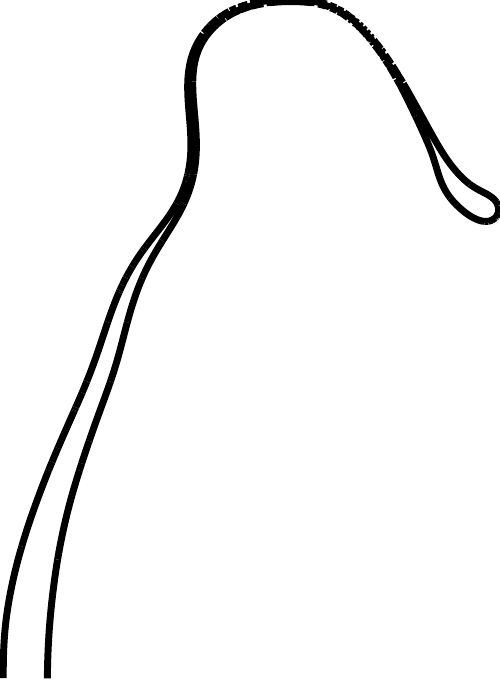}
			  } 
			& \raisebox{-0.2in}{
			 	  	\includegraphics[scale=0.3]{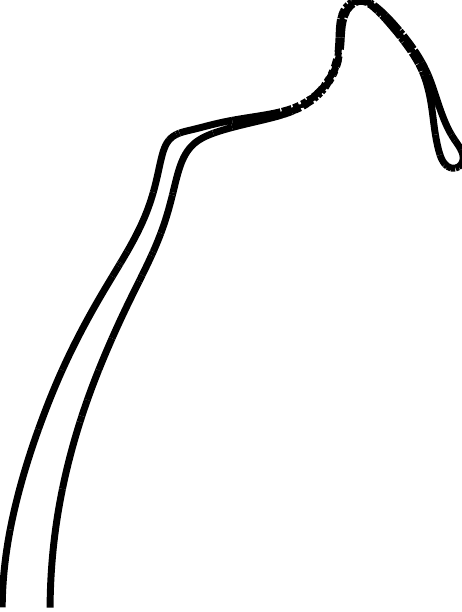}
			 	 } 
			& \raisebox{-0.2in}{
			 	 \includegraphics[scale=0.3]{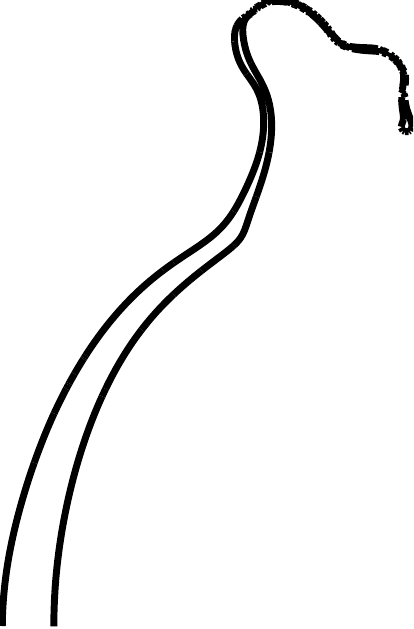}
			  } 
			& \raisebox{-0.2in}{
			 	 \includegraphics[scale=0.3]{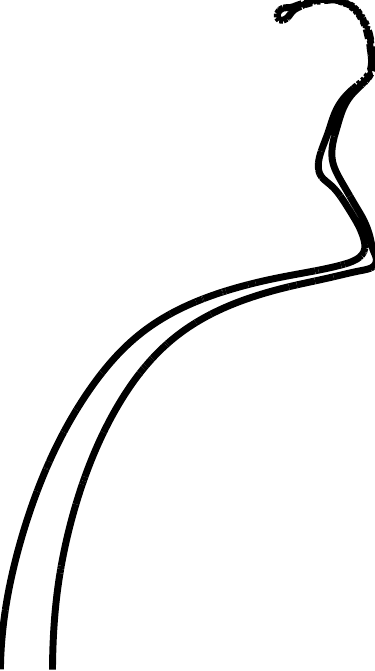}
			  } 
			& \raisebox{-0.2in}{
			 	 \includegraphics[scale=0.3]{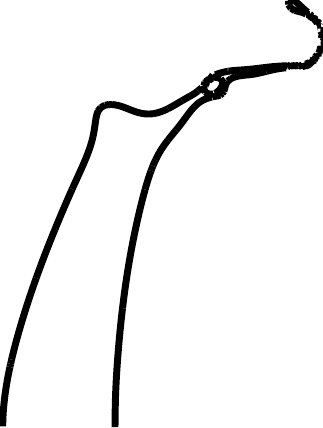}
			  }   \\ \cdashline{3-8} \\[-0.5em]

			$\textbf{$\rho^*$}$ & \textbf{200} & \raisebox{-0.2in}{
			\includegraphics[scale=0.3]{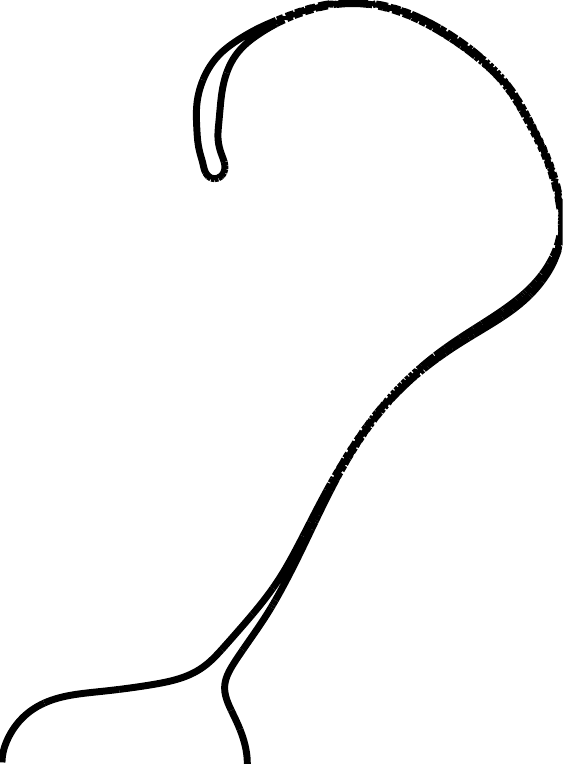}
			 	} 
			& \raisebox{-0.2in}{
			 	 \includegraphics[scale=0.3]{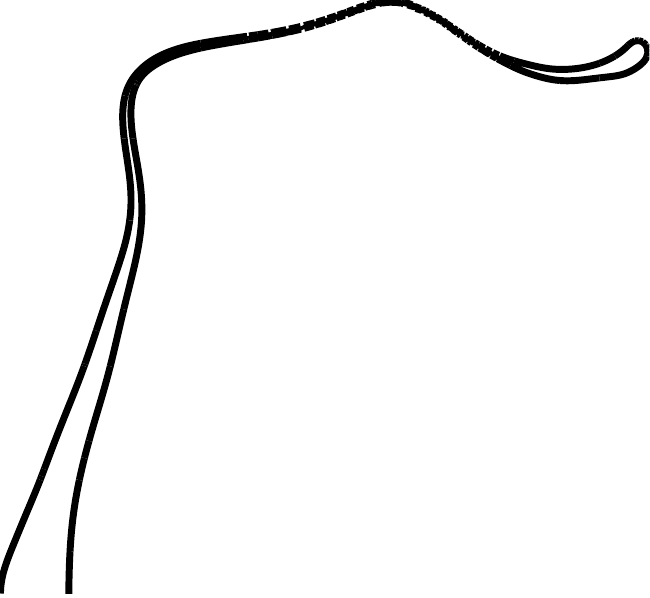}
			  } 
			& \raisebox{-0.2in}{
			 	  	\includegraphics[scale=0.3]{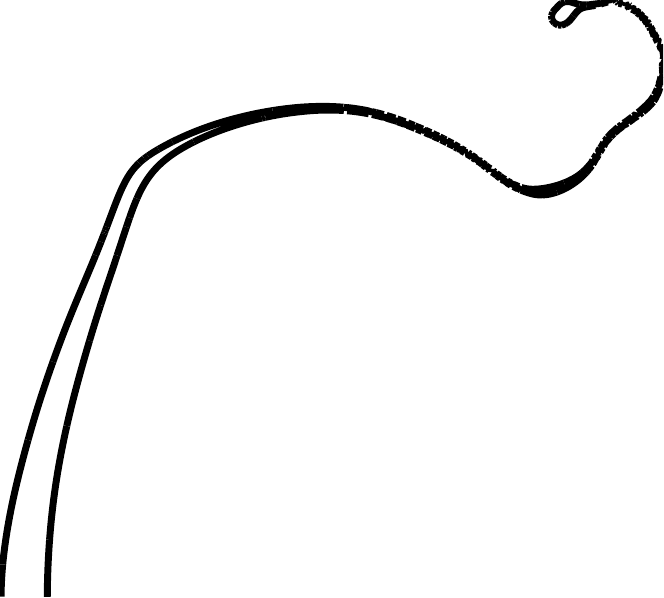}
			 	 } 
			& \raisebox{-0.2in}{
			 	 \includegraphics[scale=0.3]{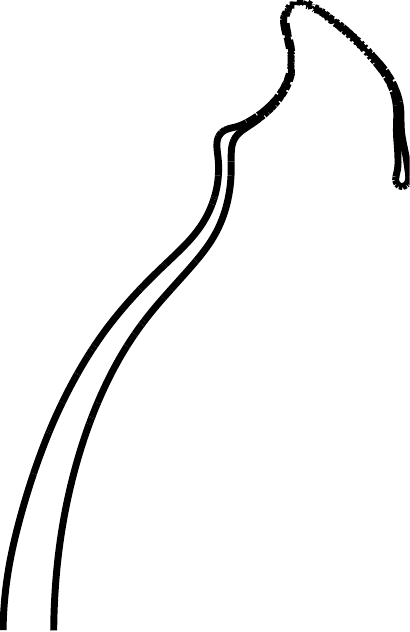}
			  } 
			& \raisebox{-0.2in}{
			 	 \includegraphics[scale=0.3]{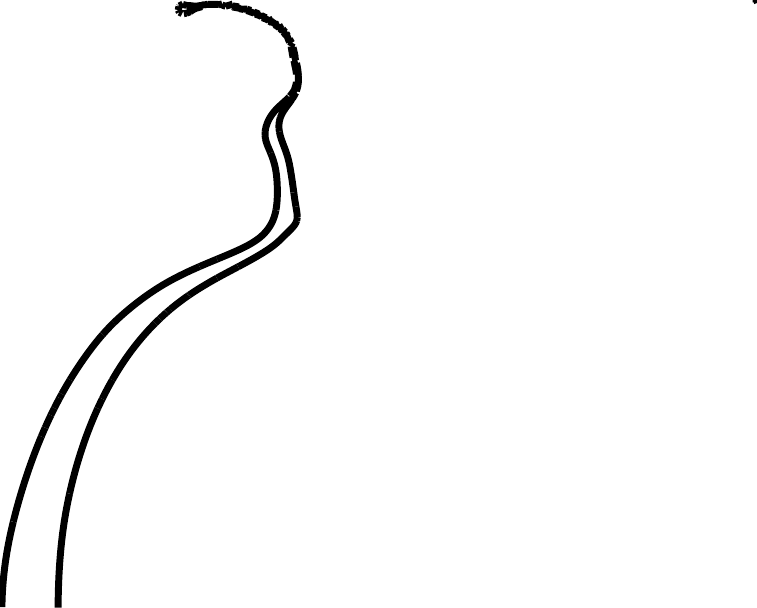}
			  } 
			& \raisebox{-0.2in}{
			 	 \includegraphics[scale=0.3]{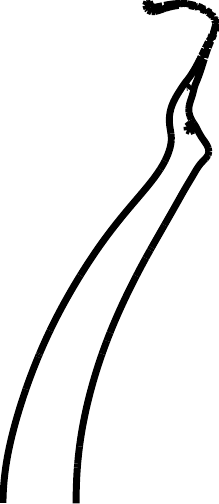}
			  }   \\ \cdashline{3-8} \\[-0.5em]

			& \textbf{250} & \raisebox{-0.2in}{
			 	\includegraphics[scale=0.3]{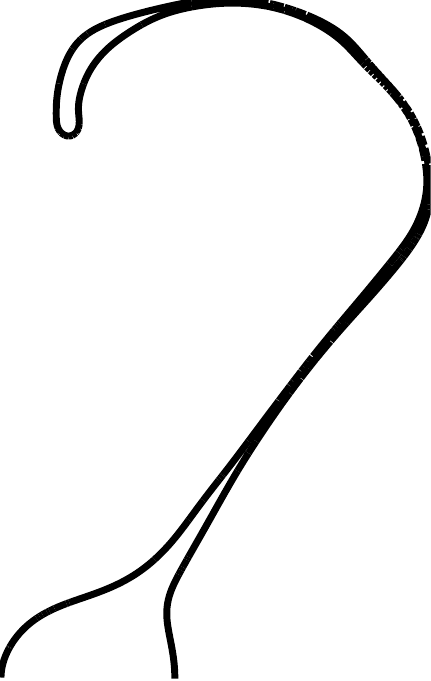}
			 	} 
			& \raisebox{-0.2in}{
			 	 \includegraphics[scale=0.3]{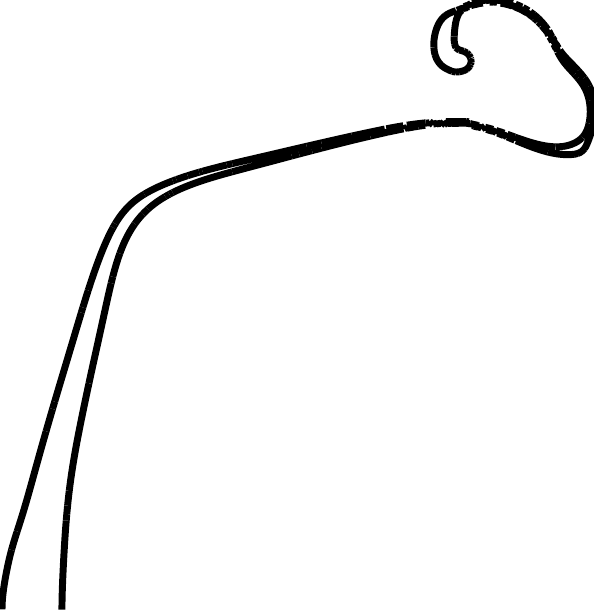}
			  } 
			& \raisebox{-0.2in}{
			 	  	\includegraphics[scale=0.3]{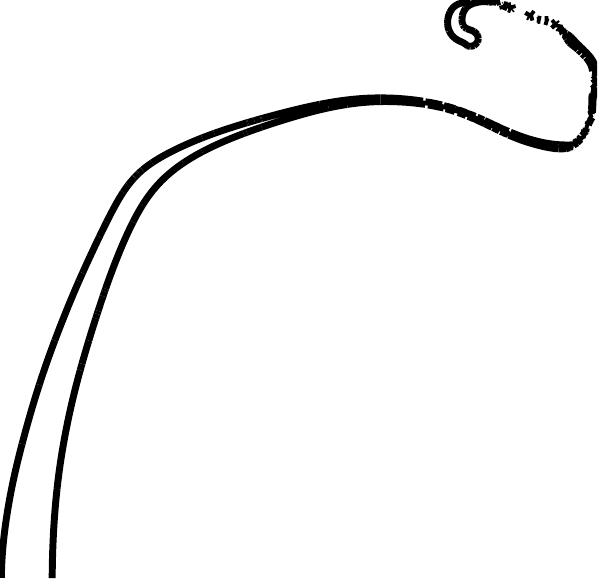}
			 	 } 
			& \raisebox{-0.2in}{
			 	 \includegraphics[scale=0.3]{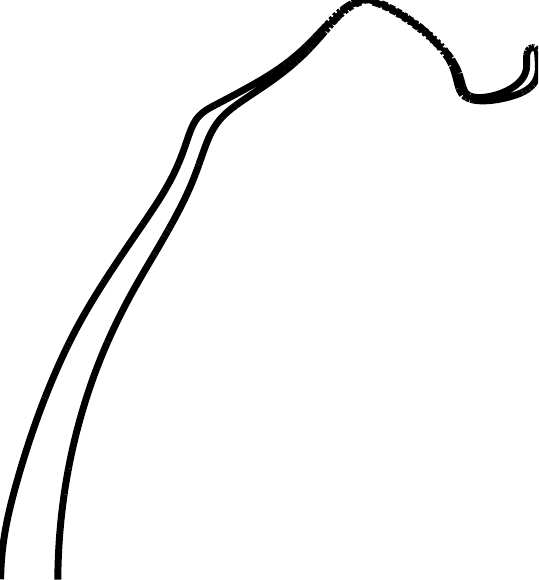}
			  } 
			& \raisebox{-0.2in}{
			 	 \includegraphics[scale=0.3]{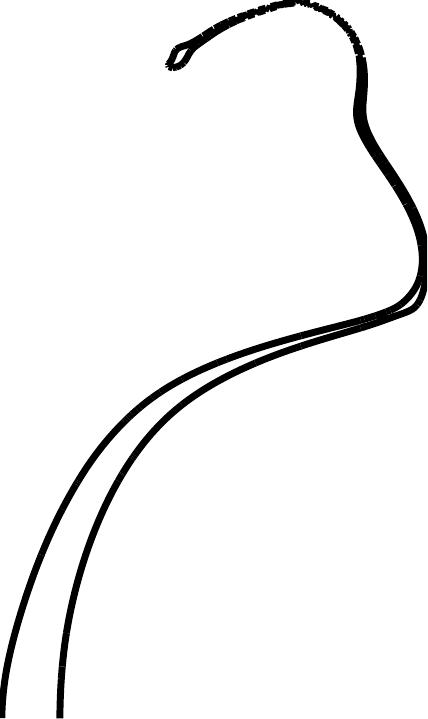}
			  } 
			& \raisebox{-0.2in}{
			 	 \includegraphics[scale=0.3]{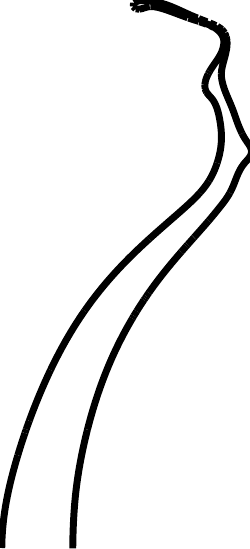}
			  }  \\ \cdashline{3-8} \\[-0.5em]

			& \textbf{500} & \raisebox{-0.2in}{
			 	\includegraphics[scale=0.35]{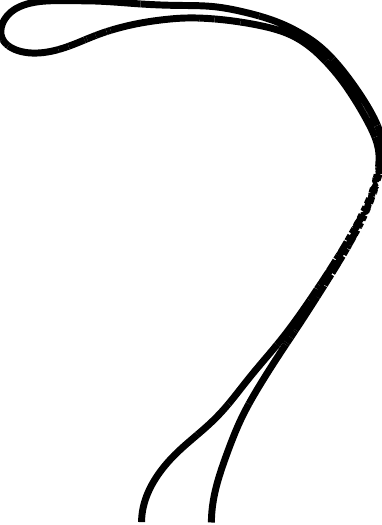}
			 	} 
			& \raisebox{-0.2in}{
			 	 \includegraphics[scale=0.3]{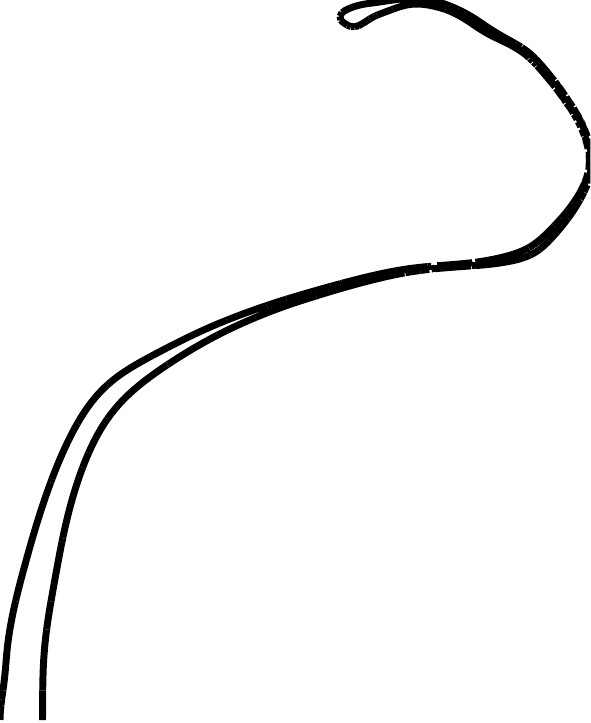}
			  } 
			& \raisebox{-0.2in}{
			 	  	\includegraphics[scale=0.3]{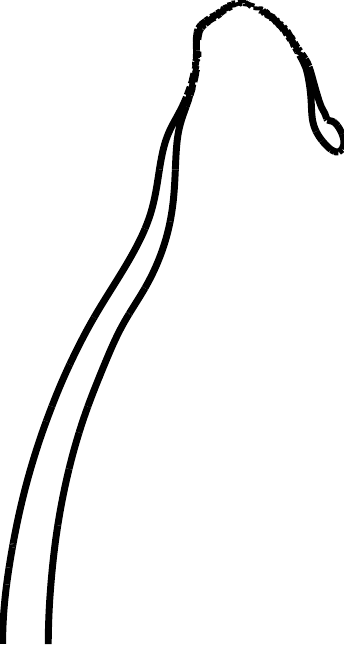}
			 	 } 
			& \raisebox{-0.2in}{
			 	 \includegraphics[scale=0.3]{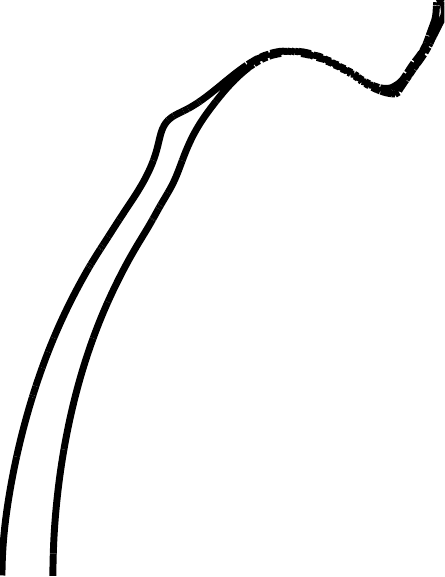}
			  } 
			& \raisebox{-0.2in}{
			 	 \includegraphics[scale=0.3]{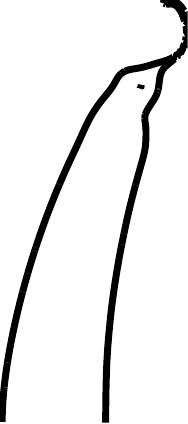}
			  } 
			& \raisebox{-0.2in}{
			 	 \includegraphics[scale=0.3]{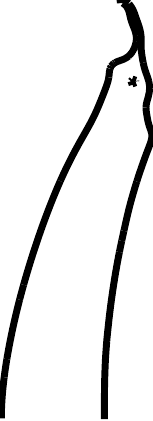}
			  } \\ \cdashline{3-8} \\[-0.5em]

			& \textbf{1000} & \raisebox{-0.2in}{
			 	\includegraphics[scale=0.3]{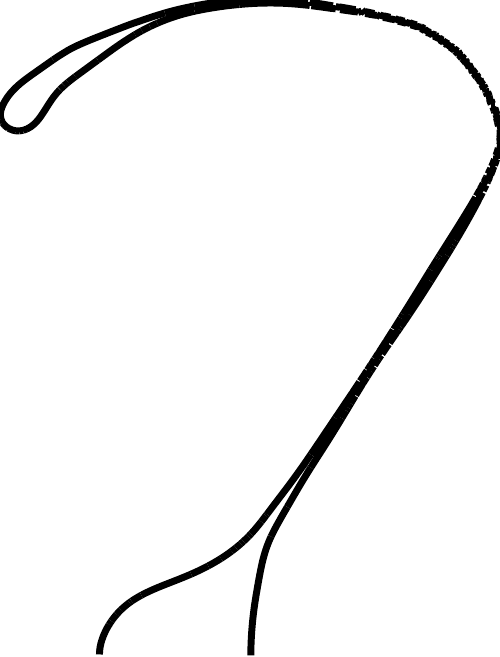}
			 	}
			& \raisebox{-0.2in}{
			 	 \includegraphics[scale=0.3]{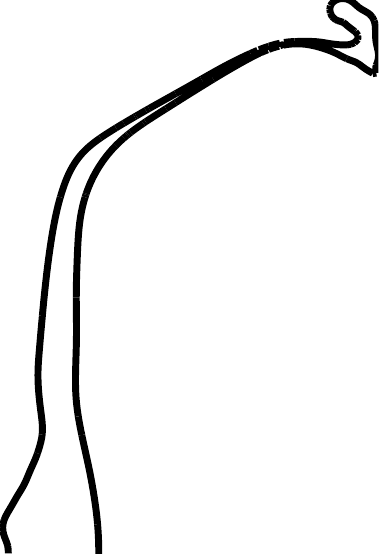}
			  } 
			& \raisebox{-0.2in}{
			 	  	\includegraphics[scale=0.3]{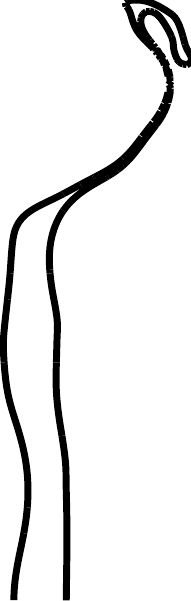}
			 	 } 
			& \raisebox{-0.2in}{
			 	 \includegraphics[scale=0.3]{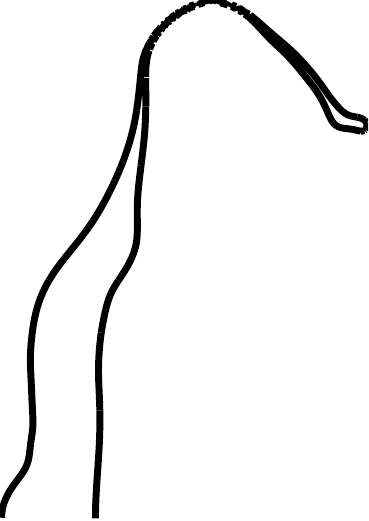}
			  } 
			& \raisebox{-0.2in}{
			 	 \includegraphics[scale=0.3]{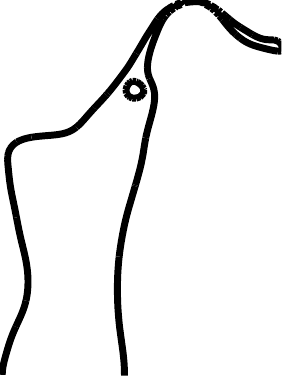}
			  } 
			& \raisebox{-0.2in}{
			 	 \includegraphics[scale=0.3]{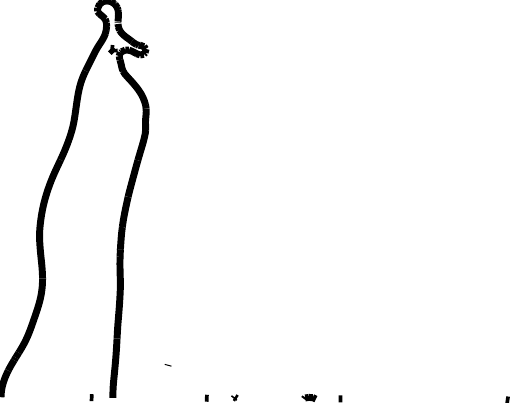}
			  }   \\ \hline \hline
		\end{tabular}%
	}
	\caption{Typical shapes of the drop at the onset of breakup for $\rho^*=10-1000$ at $We=20-120$, $Re_g=4000$, $M=100$ and $Oh=0.003-0.9$. The time $t^*_b$ where the profiles are taken is plotted in the Figure \ref{fig:time}.}
	\label{tab:char_shape}
	\end{center}
\end{table}

In order to study the morphology of the drops during breakup, typical shapes of the drops at the onset of the breakup have been tabulated in Table \ref{tab:char_shape} for all the conditions listed in the Table \ref{tab:cases}. Again, for the cases presented in the Figure \ref{fig:nobreakup}, no breakup is observed. Comparing the drop shapes for different $\rho^*$ values (for the same $We$) in the Table \ref{tab:char_shape} reveals that at $We=20$, a forward-bag (facing the gas flow) is seen for $\rho^*=10$, transient drop shapes (canopy-top which can also be seen as a shape in between forward-bag and backward-bag) for $\rho^*=50$ and $100$, and a forward-bag (bag facing the flow) with stamen for $\rho^*\geq150$. For $\rho^*=$50 and 100, the drop shapes are similar, but for $\rho^*=$50 the rim does not pinch-off from the core drop whereas, for $\rho^*=$100, the rim eventually pinches-off from the drop breaking into a toroidal ring and a smaller drop. There seems to be a progressive change with increase in $\rho^*$, from canopy shaped drop for $\rho^* = 50$ to a drop with not-so-clear stamen for $\rho^* = 100$ and very prominent stamen with a bag for $\rho^* > 150$. Interestingly, the stamen is very long for $\rho^*=150$ and it decreases in size with increase in $\rho^*$. This can be understood by studying  the velocities at the tip of the stamen. The non-dimensional velocity at the tip of the stamen (with respect to the centroid velocity of the drop), $u_{stamen}/u_{drop}=0.74$ for $\rho^*=150$, $u_{stamen}/u_{drop}=0.85$ for $\rho^*=250$ and $u_{stamen}/u_{drop}=0.9047$ for $\rho^*=1000$. This clearly implies that there is a higher relative velocity between the stamen and the drop for lower $\rho^*$ which results in more stretching of the stamen at lower $\rho^*$ values and hence results in the formation of a longer stamen. This forward-bag with stamen mode of breakup at $We=20$ was observed before in \citet{Jain2015} at $We=40$. This difference in $We$ may be due to the significantly different $Oh$ used in \citet{Jain2015} and in the present simulations, though in both the cases $Oh<0.1$ was maintained. For example, the $Oh$ used in \citet{Jain2015} was 0.1 for $\rho^*=1000$ in all the simulations, whereas here we use $Oh=0.0035$ for a similar case of $\rho^*=1000$ at $We=20$. Gas Reynolds number used for $\rho* = 1000$ discussed above is $4000$, whereas \citet{Jain2015} performed the simulations at $Re_g = 1414$. This effect of $Oh$ on the drop deformation and breakup will be discussed elsewhere in another study.


At $We=40$ and higher, a backward-bag is seen for $\rho^*=10$ (as also observed by \citet{Han2001} for $\rho^*=10$ at $Re=242$ and $We\geq37.4$), for $\rho^*=50$ a transient form of sheet-thinning, where the thin rim oscillates like a whiplash (ensuing the motion from the vortex shedding in the surrounding gas flow) and for $\rho^*=100$ and higher, drop deforms into a concave-disc (facing downstream) and eventually breaks up due to sheet-thinning. 
For $\rho^*=200-1000$ at $We = 40$ and for $\rho^*=200-250$ at $We=60$, we see an interesting ``cowboy-hat" shape of the drop. A similar drop shape was observed by \citet{Khosla2006}. For $\rho^*=10$, the length of the rim increases with an increase in $We$ value, whereas for higher $\rho^*$ ($\rho^*=$100-1000), the length of the rim decreases with increase in $We$ and at the bottom-right corner of the table for $\rho^*=500$ and 1000 at $We=100$ and 120, the drops at the onset of breakup are essentially flat discs without any discernible rim. The length and the thickness of the rim is very small that results in the formation of very fine drops during sheet-thinning breakup. Another interesting observation is that the rim is thicker for drops of lower $\rho^*$. This is possibly due to higher Taylor-Culick velocity $u_{tc}$ for lower $\rho^*$, where $u_{tc}=\sqrt{{2\sigma}/{\rho_l h}}$. Hence, a higher $u_{tc}$ and a higher stretching velocity $u_s$ at the rim (equal to the values of $u_{rim}$ in Figure 4, and is of the order of the velocity given by $\sqrt{1/\rho^*}U_g$ and acts in the opposite direction to $u_{tc}$), result in the formation of a swollen rim for the drops with lower $\rho^*$. Consequently, the stretching of the fluid in this swollen rim takes more time resulting in the delayed breakup/pinch-off of the rims of the drops for low $\rho^*$ values. This is also in agreement with our observed breakup time, $t^*_b$, for $\rho^*=10$ case as shown later in the Figure \ref{fig:time}.



\begin{figure}
\centering
\includegraphics[width=4.in]{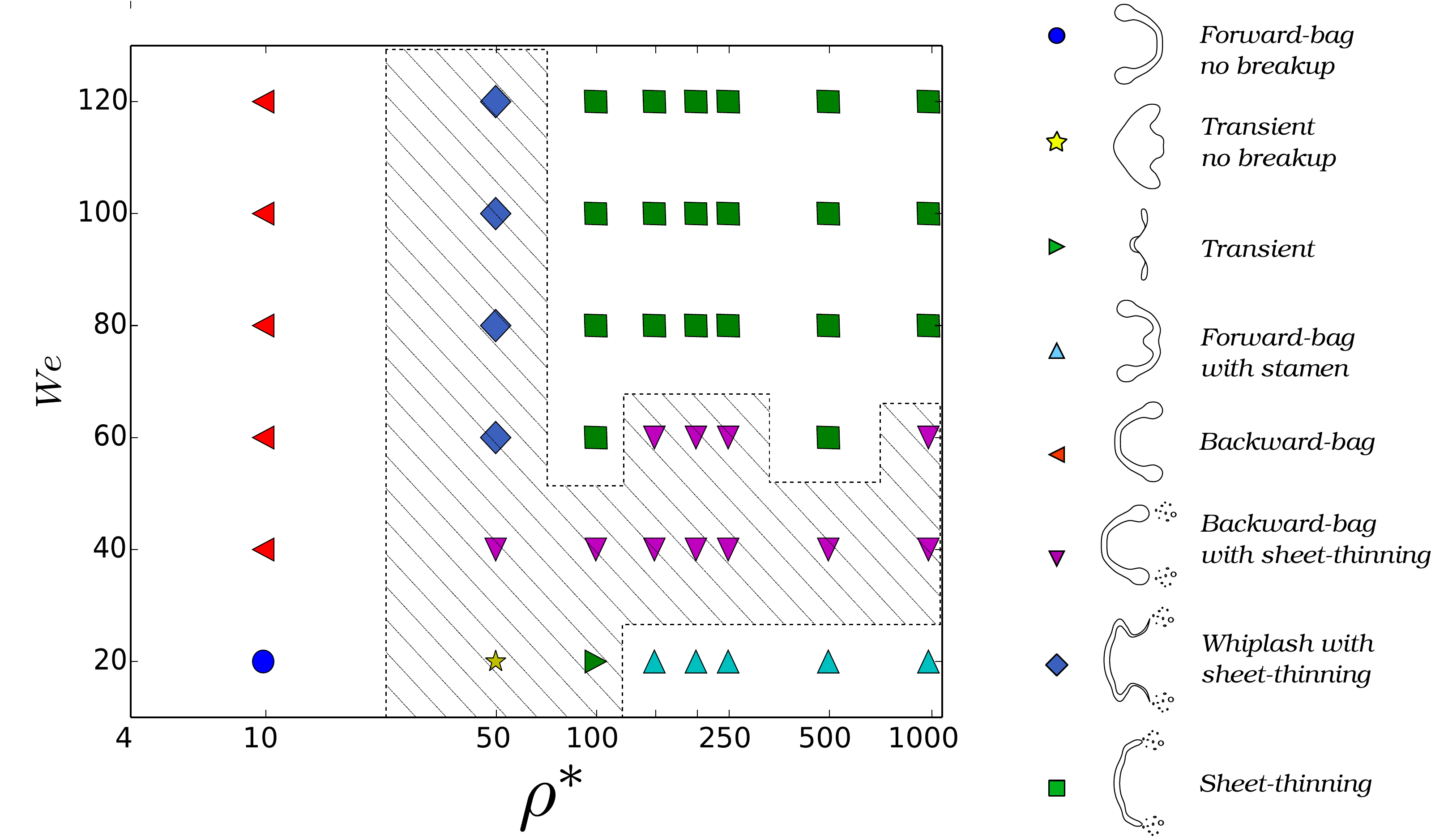}
\centering
\caption{ Phase plot $\rho^*-We$ along with the typical drop shapes for breakup modes shown on the right. Hatched region shows the transition regime. The $\rho^*-$ axis is scaled to the logarithm base of 2.}
\label{fig:phaseplot}
\end{figure}

\begin{figure}[t!]
    \centering
    \begin{subfigure}[t]{\textwidth}
        \centering
        \includegraphics[width=\textwidth]{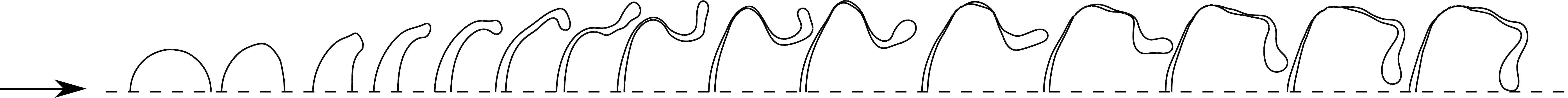}
        \caption{$\rho^*=10, We=60$}
        \label{fig:ev_dr10we60}
    \end{subfigure}
    ~ 
    \begin{subfigure}[t]{\textwidth}
        \centering
        \includegraphics[width=\textwidth]{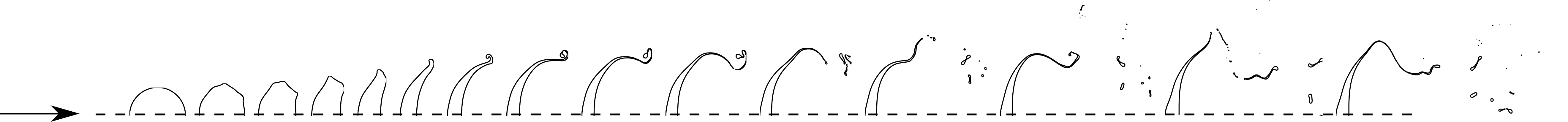}
        \caption{$\rho^*=50, We=60$}
        \label{fig:ev_dr50we60}
    \end{subfigure}
    ~
    \begin{subfigure}[t]{\textwidth}
        \centering
        \includegraphics[width=\textwidth]{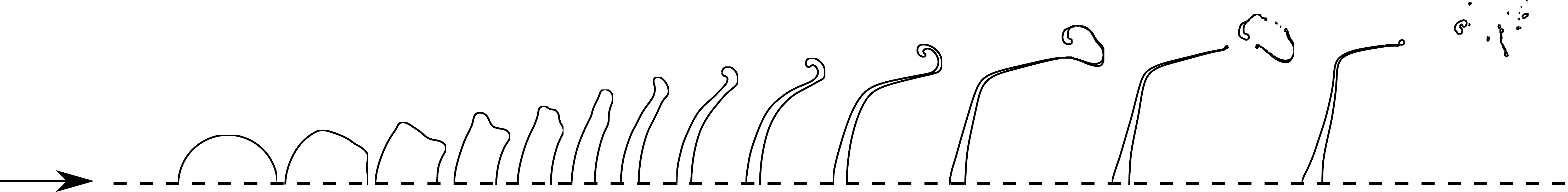}
        \caption{$\rho^*=250, We=40$}
        \label{fig:ev_dr250we40}
    \end{subfigure}
    ~
    \begin{subfigure}[t]{\textwidth}
        \centering
        \includegraphics[width=\textwidth]{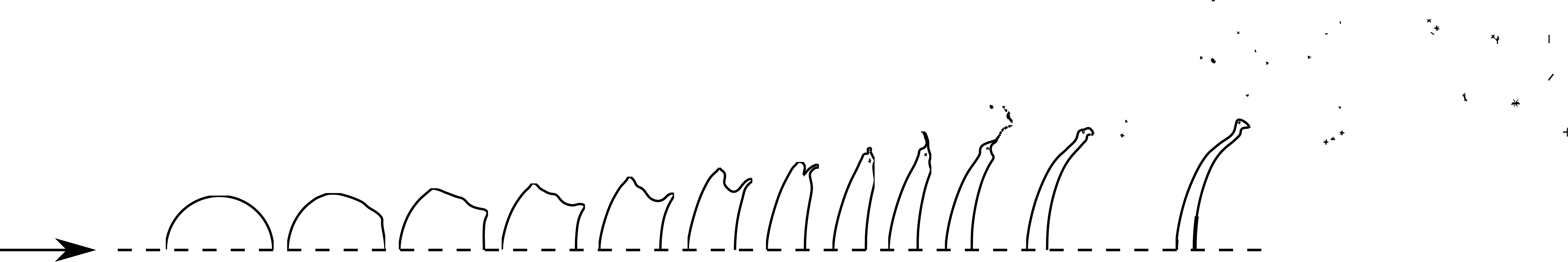}
        \caption{$\rho^*=500, We=120$}
        \label{fig:ev_dr500we120}
    \end{subfigure}
    \caption{Evolution of the drop shape in time for $\rho^*=10$ at $We$=60, $\rho^*=50$ at $We$=60, $\rho^*=250$ at $We$=40 and $\rho^*=500$ at $We=120$. Arrows show the direction of gas flow and the dotted lines mark the axis of symmetry. Note that the distance between consecutive droplet profiles plotted here does not represent the actual displacement of the drop.}
    \label{fig:shape_class}
\end{figure}

\begin{figure}[t!]
    \centering
    \begin{subfigure}[c]{0.5\textwidth}
        \centering
        \includegraphics[width=\textwidth]{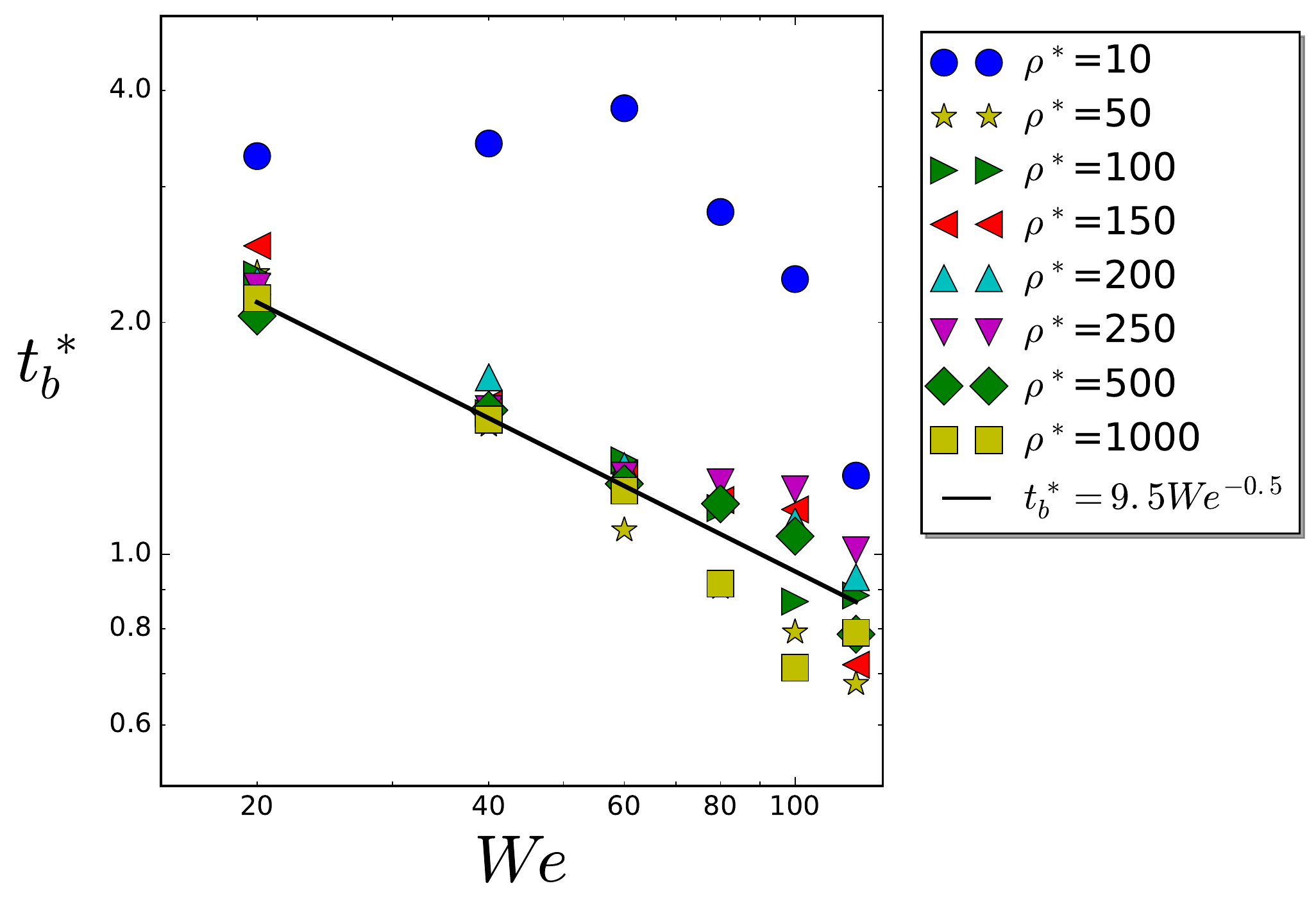}
        \caption{Time at the onset of breakup. Solid line represents the power law fit.}
        \label{fig:time}
    \end{subfigure}%
    ~ 
    \begin{subfigure}[c]{0.45\textwidth}
        \centering
        \includegraphics[width=\textwidth]{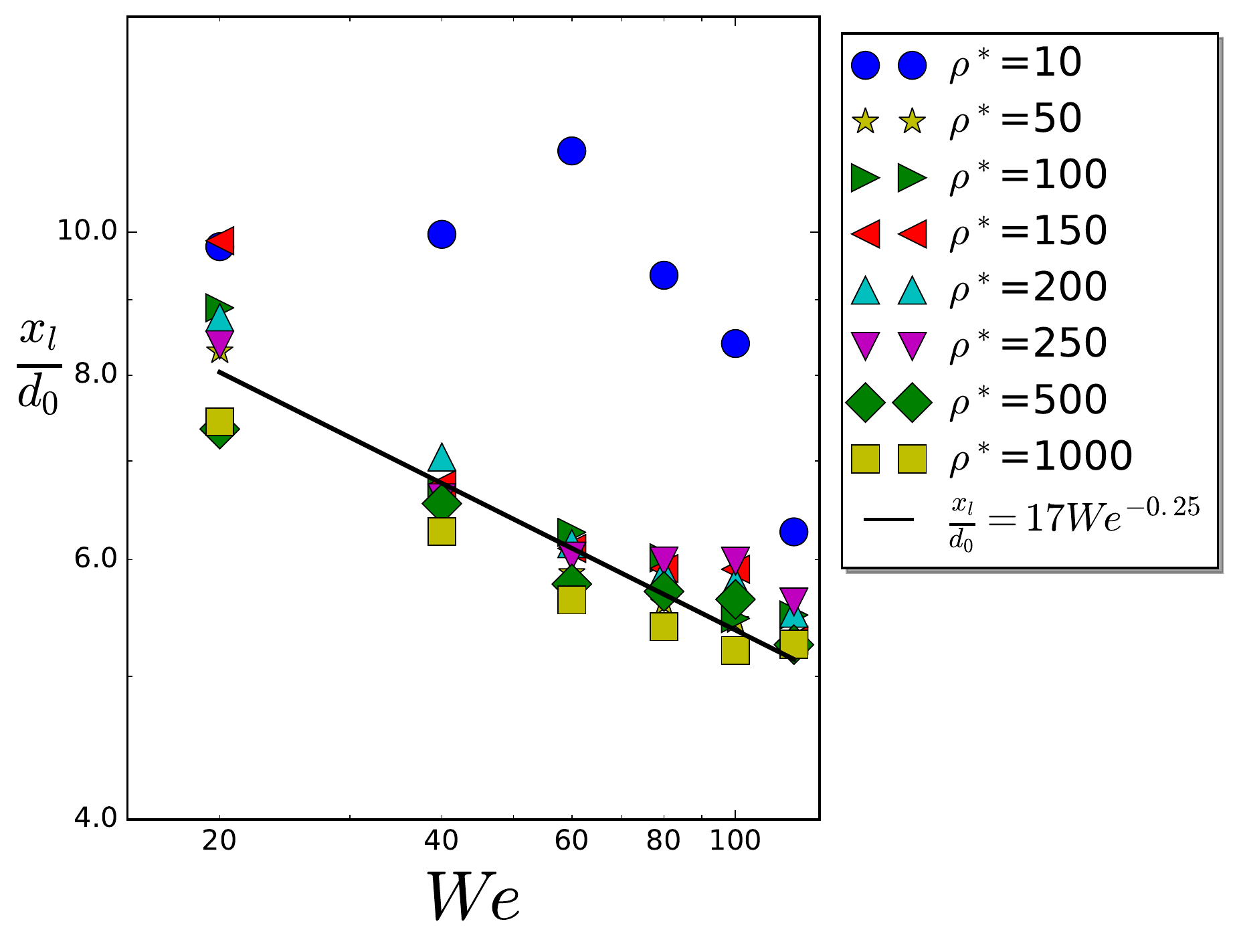}
        \caption{Distance travelled by the drop up to the onset of breakup. Solid line represents the power law fit.}
        \label{fig:dist}
    \end{subfigure}
    ~
    \begin{subfigure}[c]{0.5\textwidth}
        \centering
        \includegraphics[width=\textwidth]{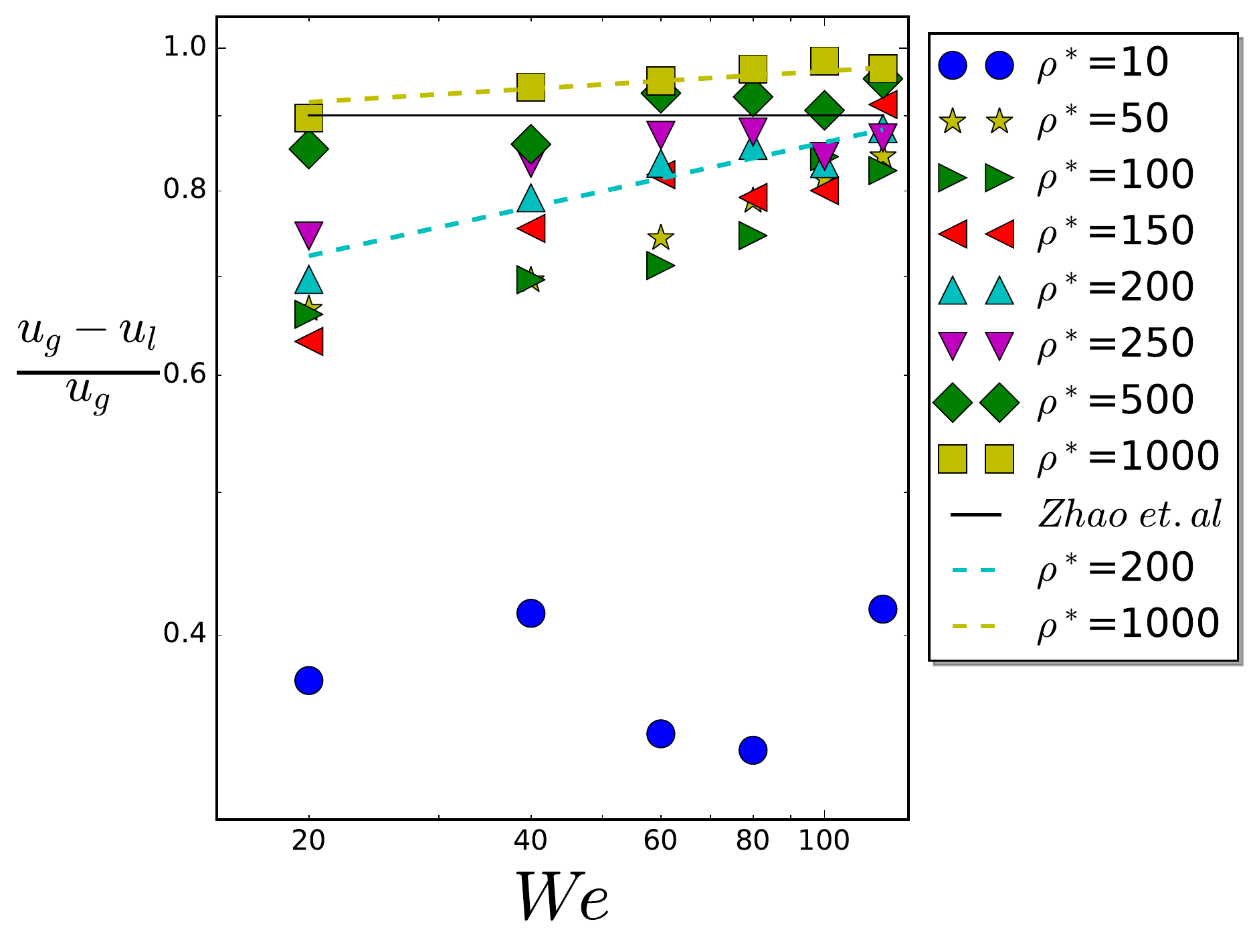}
        \caption{Relative velocity at the onset of breakup. Solid line represents the results by \citet{Zhao2010}. Dotted lines shows the general power law fit for $\rho^*=1000$ and $\rho^*=200$ case.}
        \label{fig:rel_vel}
    \end{subfigure}
    \caption{Relative velocities, time taken and the distance travelled by the drop at the onset of breakup for the parameters listed in the Table \ref{tab:char_shape}.}
    \label{fig:onset}
\end{figure}

To summarize the breakup modes presented in the Table \ref{tab:char_shape}, we draw a phase plot of $\rho^*$ vs $We$ shown in the Figure \ref{fig:phaseplot}. Typical shapes for each breakup mode is shown beside the plot. We characterize the drop shapes as following, from top to bottom: Forward-bag no-breakup, Transient no-breakup, Transient, Forward-bag with stamen, Backward-bag, Backward-bag with sheet-thinning, whiplash with sheet-thinning and finally the bottom most is sheet-thinning. Hatched region marks the transition regime indicating transition from bag(forward/backward) to sheet-thinning. Note, the differences between the drop shapes for Backward-bag with sheet thinning (for E.g. $\rho^*=250$ at $We=40$), whiplash with sheet-thinning (for E.g. $\rho^*=50$ at $We=60$) and sheet-thinning (for E.g. $\rho^*=500$ at $We=120$) are not very evident from the instantaneous shapes of the drop presented in Table \ref{tab:char_shape} at the onset of breakup. However, the temporal evolution of these drop shapes (see Figure \ref{fig:shape_class}) suggest the classification shown in Figure \ref{fig:phaseplot}. Figure \ref{fig:ev_dr10we60} shows the time evolution of the drop for $\rho* = 10$ and $We = 60$. The whiplash of the rim of the drop for $\rho* = 10$ is clearly very different from that for $\rho* = 50$ shown in Figure \ref{fig:ev_dr50we60}. For $\rho* = 50$ and $We = 60$, time evolution in figure \ref{fig:ev_dr50we60} shows the whiplash action of the drop rim along with the sheet-thinning at the edges of the drop. Figure \ref{fig:ev_dr250we40}, for $\rho* = 250$ and $We = 40$, shows the formation of backward bag and detachment of its rim. Sheet-thinning sets up only after the rim detaches. Figure \ref{fig:ev_dr500we120}, for $\rho* = 500$ and $We = 120$, shows the sheet-thinning mode of breakup. To make the differences between these breakup modes and the backward bag even more clear, Figure \ref{fig:ev_dr10we60}, for $\rho* = 10$ and $We = 60$, shows the time evolution of the drop for the backward bag case.

In addition to these differences in the deformation, breakup morphologies and breakup modes, the breakup mechanism is also different for higher and lower $\rho^*$ values. Breakup is due to the RT instability at higher $\rho^*$ values ($\rho^*\ge150$) \citep{Zhao2010,Jain2015}, whereas breakup is due to the dynamics of the rim at lower $\rho^*$ values and is significantly influenced by the surrounding gas flow (Section  \ref{sec:rim}). Hence, drops for roughly $\rho^*\ge150$ behave similarly at similar values of $We$. This difference in breakup for different $\rho^*$ values (with $Re_g$, $M$ being constant and $Oh<0.1$) makes ``\textit{Density Ratio}" a crucial parameter in characterizing the secondary breakup of drops.


Figures \ref{fig:time}, \ref{fig:dist} and \ref{fig:rel_vel} show the drop breakup time $t^*_b$, the drop displacement, $x_l/d_0$, and the relative velocity of the centroid of the drop, $u_r=(U_g-u_l)/(U_g)$, respectively, at the onset of breakup for the conditions listed in the first row of the Table \ref{tab:char_shape}. Clearly, $t^*_b$ and $x_l/d_0$ are quite different for the drops with $\rho^*=10$ and for the drops with $\rho^*=50-1000$. With an increase in $We$, both $t^*_b$ and $x_l/d_0$ decrease following a power-law given by $t_b^*=9.5We^{-0.5}$ and $x_l/d_0 = 17We^{-0.25}$, respectively. Relative velocity $u_r$ on the other hand has a continuous variation from $\rho^*=50$ to $\rho^*=1000$ following a general power-law given by $4(10^{-4}\rho + 0.1)We^{(0.13 - 10^{-4}\rho)}$ with average values increasing from 0.76 to 0.95, though it is significantly different for $\rho^*=10$ with an average value of 0.36.  \citet{Zhao2010} reported an average value of 0.9 for ethanol and water drops combined, which are in good agreement with the simulations presented here (also shown in the figure as a line) and \citet{Dai2001} reported 0.87 for water drops. Relative velocity, $u_r$, increases with an increase in $\rho^*$ value indicating that the drops for lower $\rho^*$ would attain higher velocity at the onset of breakup. Here, we note that the drops for $\rho^*=10$ at $We=20$ and 40 and for $\rho^*=50$ at $We=20$, do not breakup at all. This is in good agreement with the observations of \citet{Han2001}. The values corresponding to these values of $\rho^*$ and $We$ reported in the Figure \ref{fig:onset} indicate only a tendency to breakup. This tendency to breakup is based on the criteria that the drop could have pinched-off at the thinnest section attained during the deformation process. However, when the simulations are run for a longer duration, the rim retracts and the breakup does not occur.

\subsection{Rayleigh-Taylor instability} \label{sec:rt}



Role of Rayleigh-Taylor instability in the breakup of a drop in the catastrophic regime has been extensively discussed in the last decades \citep{Harper1972,Simpkins1972,Joseph1999,Patel1981, Lee2000, Guildenbecher2009}. \cite{Harper1972} showed that, at high Bond numbers (above $10^5$), Rayleigh-Taylor instability dominates the algebraic aerodynamic deformation and leads to formation of waves on the windward side of drops.
However, bag formation is mostly considered as blowing out of a thin liquid sheet due to large stagnant pressure. Nevertheless, formation of stamen and multibag at higher Weber number suggest role of instability in the formation of bag with a decrease in wavelength with an increase in the acceleration of the droplet. \cite{Theofanous2004} showed that the Rayleigh-Taylor based theoretical predictions of number of waves agreed well with the experimental observations for drops with a wide range of viscosity values. Experimental observations of \cite{Zhao2010} also showed good agreement with the theoretical expression for the number of expected waves (1 for bag and 2 for bag-with-stamen) proposed by \cite{Theofanous2004}. 

\begin{figure}
\centering
\includegraphics[width = 4.in]{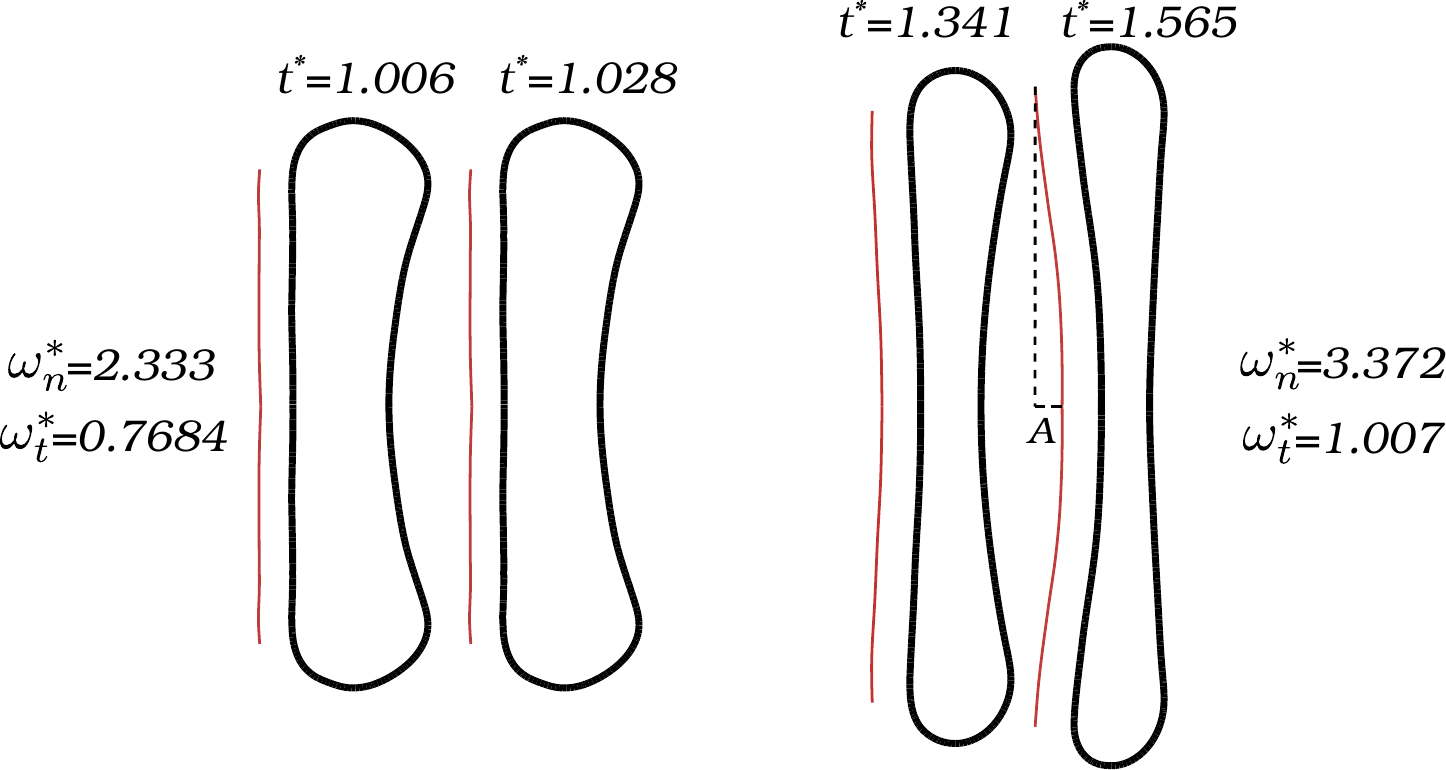}
\centering
\caption{Representation of RT waves on the drop for $\rho^*=1000$ at $Re_g=1414$, $M=1000$, $We=20$ and at different $t^*$  values mentioned in the figure. Red thin-lines beside the drop represent the RT waves on the windward surface of the drop.}
\label{fig:waves}
\end{figure}

In our numerical simulations, we also observe the formation and growth of RT waves on the windward surface of the drop as shown in the Figure \ref{fig:waves}. Shapes of the drop for $\rho^*=1000$ at $Re_g=1414$, $M=1000$, $We=20$ is shown at 4 different times, $t^*=1.006,1.028,1.341$ and $1.565$. It is quite evident from these figures that the amplitude, $A$, of the wave grows with time. The drop eventually deforms into a forward-bag and then breaks. We assume this wave on the surface of the drop as an RT wave (highlighted in the Figure \ref{fig:waves} using thin red lines) and calculate the non-dimensional RT wavenumber in the maximum cross-stream direction of the drop at $t^*\sim1.3$ (breakup initiation time at $We=20$ \citep{Xiao2014}) as,   

\begin{equation}
    N_{RT}=\frac{D_{max}}{\lambda_{max}}=\frac{0.1754}{0.18}=0.97
\end{equation}

This value of $N_{RT}=0.97$ lies in the range of $1/\sqrt{3} < N_{RT} < 1.0$, which implies that the drop deforms into a forward-bag \citep{Zhao2010} and breaks. This is in very good agreement with the observations in our simulations. Growth rate of the RT wave obtained from the numerical simulations, $\omega_n$, can be calculated by comparing the amplitude of the wave, $A$ (shown in the Figure \ref{fig:waves}) at two different time instances; assuming normal mode growth of the waves, $A=A_0e^{\omega_n t}$. Considering the surface tension effects to be negligible, theoretical growth rate, $\omega_t$, can be calculated using the following expression,

\begin{equation}
    \omega_t=\sqrt{k a \bigg(\frac{\rho_l-\rho_g}{\rho_l+\rho_g}\bigg)}.
    \label{eq:growthrate}
\end{equation}

We calculated the growth rate of these RT waves at two time instances (a)$t^*_1=1.006$ to $t^*_2=1.028$ and (b)$t^*_1=1.341$ to $t^*_2=1.565$. Note that the acceleration of the drop changes over time due to the change in the drag force with the change in shape and velocity. Therefore, the corresponding growth rate of the instability is expected to change. Nevertheless, we verify here that the dependency of the growth rate of the instability on the instantaneous Bond number ($Bo = \rho_l a d_0^2/4\sigma$) remains unchanged (see the relation in Appendix \ref{app:growthrate}). For the case (a) non-dimensional numerical growth rate is found to be 
$\omega^*_n=2.31 {Bo^{\frac{3}{4}}}/{\sqrt{We}}$ and non-dimensional theoretical growth rate is found to be 
$\omega^*_t=0.7598 {Bo^{\frac{3}{4}}}/{\sqrt{We}}$, where the instantaneous Bond number is $7.48$, and for case (b) 
$\omega^*_n=2.54 {Bo^{\frac{3}{4}}}/{\sqrt{We}}$, where the instantaneous Bond number is $10.73$. Interestingly, for both the cases, $\omega^*_n$ is $\sim3$ times that of $\omega^*_t$. These significantly different, yet consistent growth rate values, could be due to the end-effects associated with the growth of RT waves on the surface of the drop.

\subsection{Flow around the drop and the effect of Reynolds number} \label{sec:flow}

\begin{figure}
    \centering
    \begin{subfigure}[c]{0.4\textwidth}
        \centering
        \includegraphics[width=\textwidth]{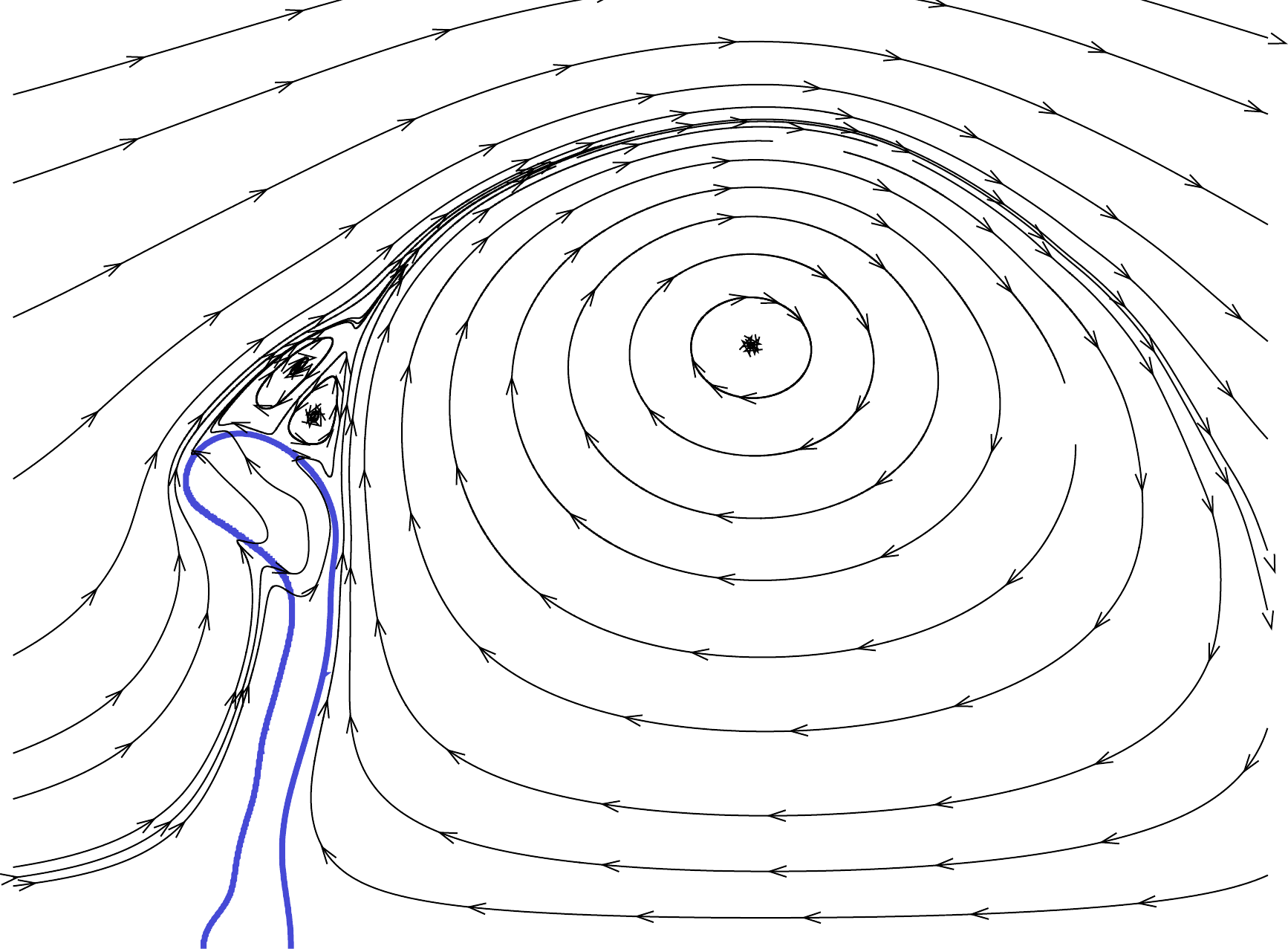}
        \caption{2D axisymmetric simulation. Blue line represents the drop and lines with the arrows are the streamlines of the velocity relative to the centroid velocity of the drop.}
        \label{fig:flow_2d_re4000}
    \end{subfigure}%
    ~ 
    \begin{subfigure}[c]{0.6\textwidth}
        \centering
        \includegraphics[width=\textwidth]{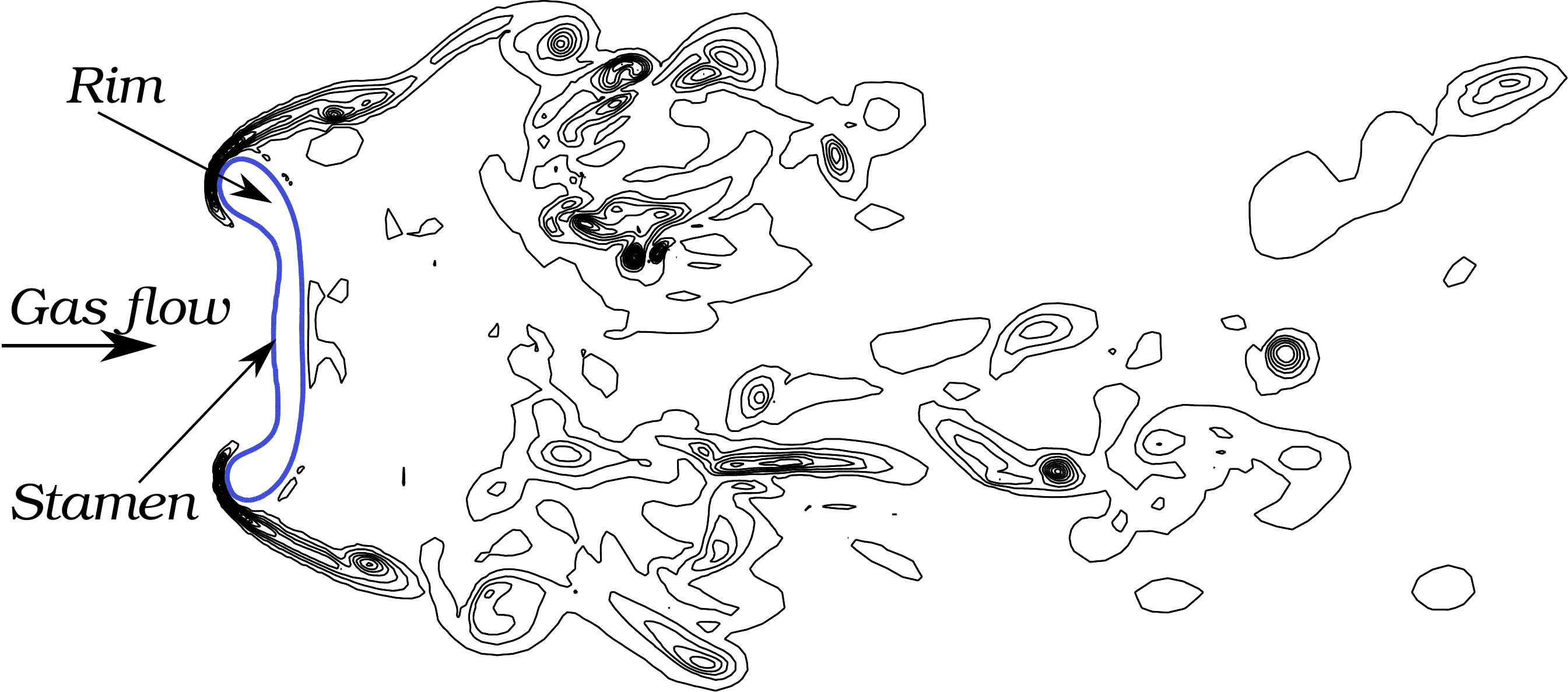}
        \caption{Cross-section view of the 3D simulation. Blue line represents the drop, black lines represents the vorticity contour plot. Arrow represents the direction of gas flow.}
        \label{fig:flow_3d_re4000}
    \end{subfigure}
    ~
    \begin{subfigure}[c]{0.6\textwidth}
        \centering
        \includegraphics[width=\textwidth]{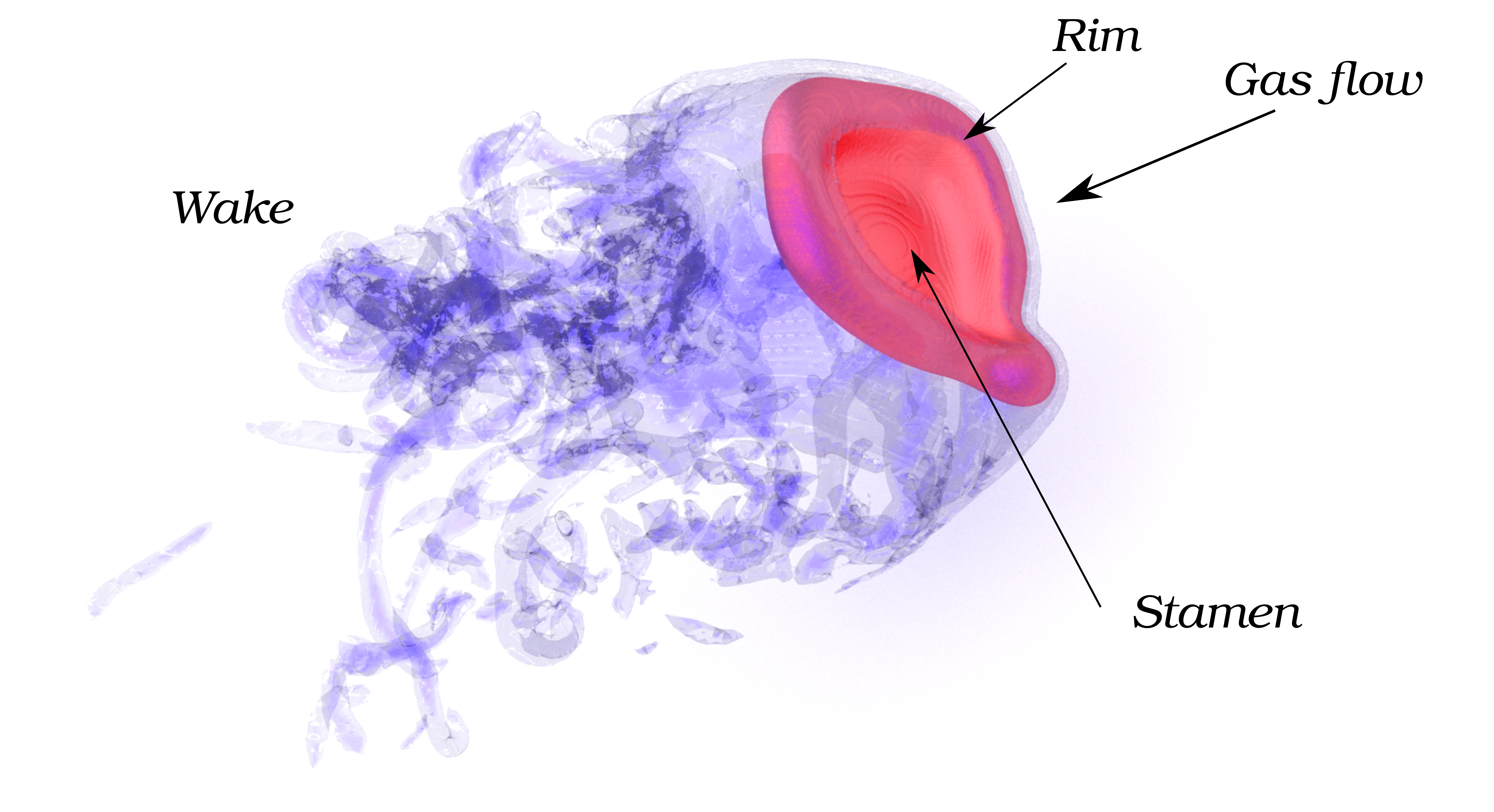}
        \caption{3D simulation in rendered view, showing the drop and the vorticity isosurfaces. Arrow represents the direction of gas flow.}
        \label{fig:render_re4000}
    \end{subfigure}
    \caption{Comparison of flow field around the drop for $\rho^*=1000$ at $Re_g=4000$, $M=1000$, $We=20$ and at $t^*=1.52$ obtained from a 3D and 2D axisymmetric simulation.}
    \label{fig:flow_re4000}
\end{figure}

\begin{figure}
    \centering
    \begin{subfigure}[c]{0.4\textwidth}
        \centering
        \includegraphics[width=\textwidth]{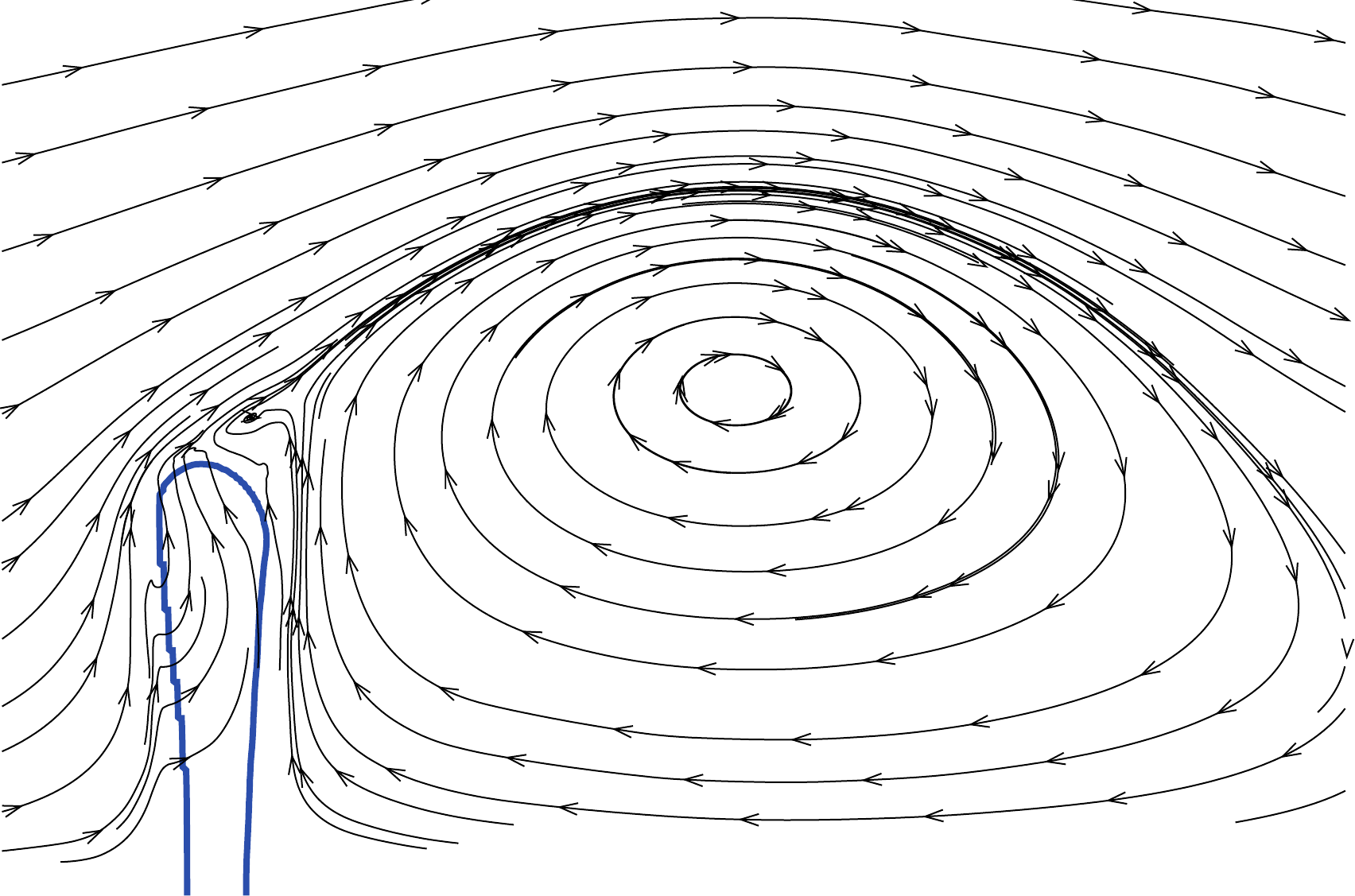}
        \caption{2D axisymmetric simulation. Blue line represents the drop and lines with the arrows are the streamlines of the velocity relative to the centroid velocity of the drop.}
        \label{fig:flow_2d_re1414}
    \end{subfigure}%
    ~ 
    \begin{subfigure}[c]{0.6\textwidth}
        \centering
        \includegraphics[width=\textwidth]{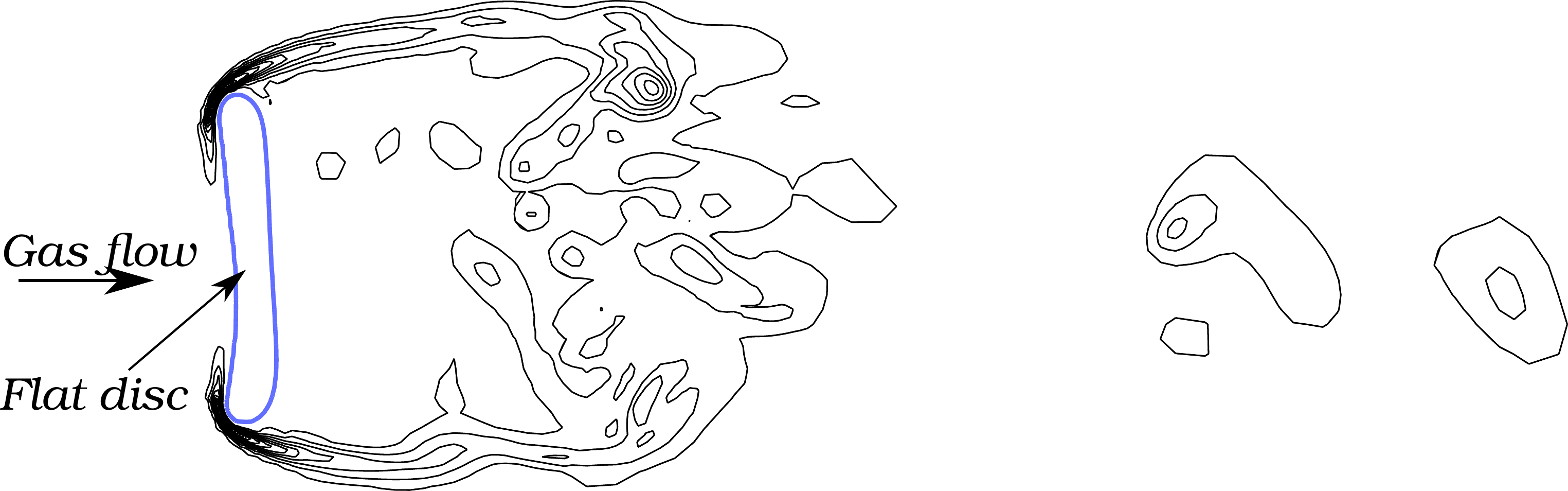}
        \caption{Cross-section view of the 3D simulation. Blue line represents the drop, black lines represents the vorticity contour plot. Arrow represents the direction of gas flow.}
        \label{fig:flow_3d_re1414}
    \end{subfigure}
    ~
     \begin{subfigure}[c]{0.6\textwidth}
        \centering
        \includegraphics[width=\textwidth]{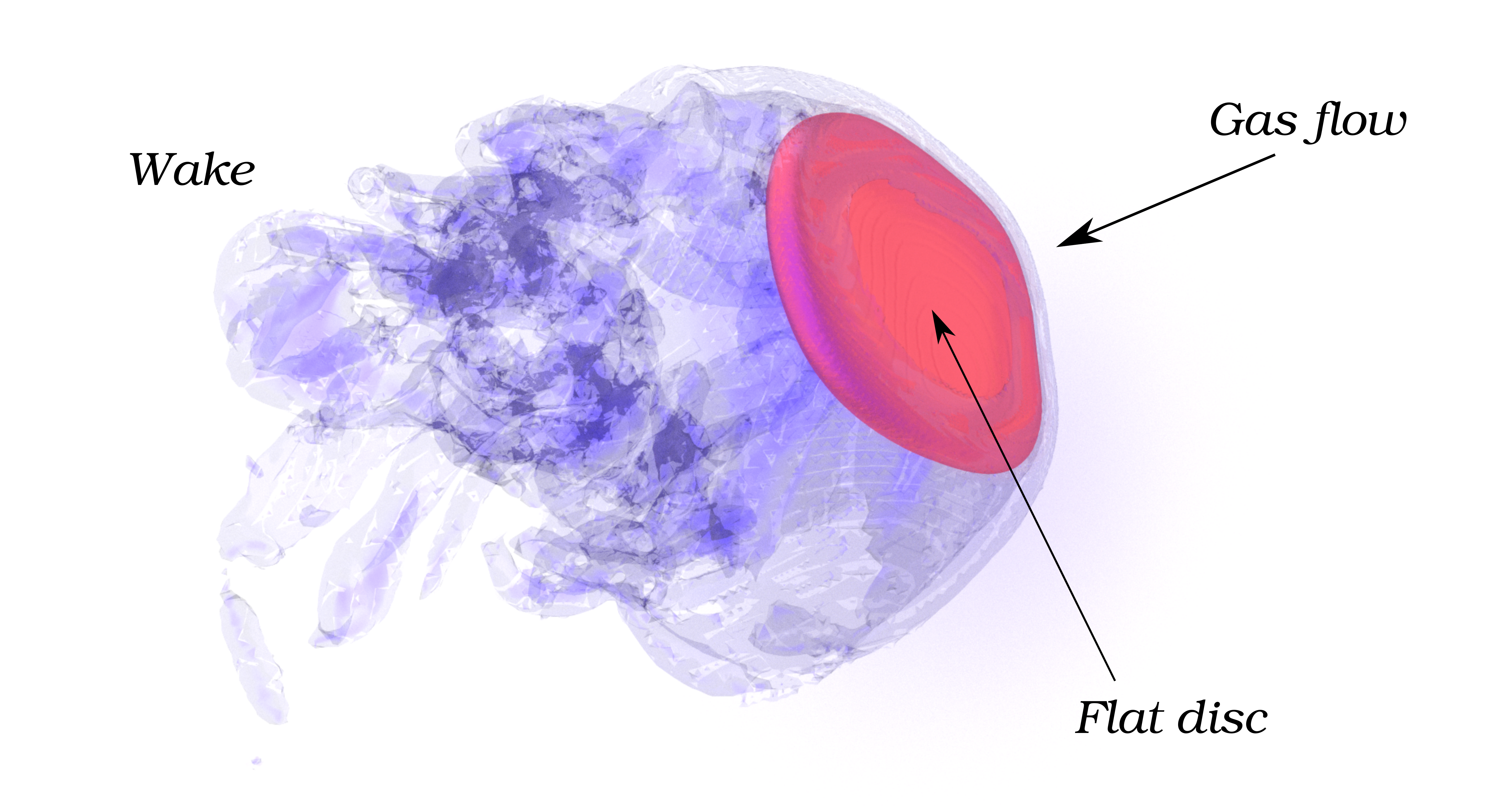}
        \caption{3D simulation in rendered view, showing the drop and the vorticity isosurfaces. Arrow represents the direction of gas flow.}
        \label{fig:render_re1414}
    \end{subfigure}
    \caption{Comparison of flow field around the drop for $\rho^*=1000$ at $Re_g=1414$, $M=1000$, $We=20$ and at $t^*=1.52$ obtained from a 3D and 2D axisymmetric simulation.}
    \label{fig:flow_re1414}
\end{figure}

In this section, we study the flow field around the drop to identify the effects of the flow structures on the deformation and breakup of the drops. Although, some attempt has been made to comprehend this through experimental and numerical observations (see \cite{Flock2012, Strotos2016}), a more systematic analysis is required to completely characterize the flow field. Figures \ref{fig:flow_re4000} and \ref{fig:flow_re1414} show the flow field around the drop at $t^*=1.52$ for $\rho^*=1000$, $M=1000$ and $We=20$ at two different gas Reynolds numbers, $Re_g=4000$ and $1414$, obtained from 3D and 2D axisymmetric simulations. We observe the formation of a Hill's spherical vortex in the wake region for the drop in axisymmetric simulations (Figures \ref{fig:flow_2d_re4000} and \ref{fig:flow_2d_re1414}), whereas in 3D simulations (Figures \ref{fig:flow_3d_re4000} and \ref{fig:flow_3d_re1414}), a three-dimensional vortex shedding is seen in the wake region for both the cases. At very early stages of the drop deformation we also observed the formation of vortex ring in the 3D simulations (not shown here). This vortex ring starts to shed at around $t^*\sim0.15$ for the drop with $\rho^*=1000$ value and subsequently at $t^*=1.52$, we see a three-dimensional eddy formation and stretching resulting in a highly unsteady complex flow in the wake region of the drop. Though, the formation of a vortex and its shedding behind the drop need not be similar to the flow past a cylinder, a qualitative comparison could help us in better understanding the flow. \citet{Jeon2004} performed a comparative study of the the vortex in the wake of the cylinder and the formation of a vortex ring and reported that the shedding of the vortex starts at  the non-dimensional time, $t/t_{cv}=4$, expressed in terms of the characteristic time $t_{cv}=d_0/\overline{U}$, where $\overline{U}$ is the average velocity of the vortex generator (relative velocity of the drop with respect to gas, $U_g-u_{drop}$ in our case) and $d_0$ is its diameter. For the present case, we find that the relative velocity, $(U_g-u_{drop})/U_g$ is 0.92 from our simulations. Hence the vortex shedding is expected to start at $t^*=0.14$, which is in very good agreement with our observations of $t^*\sim0.15$ in our simulations.

\begin{figure}
\centering
\includegraphics[height=0.3\textheight]{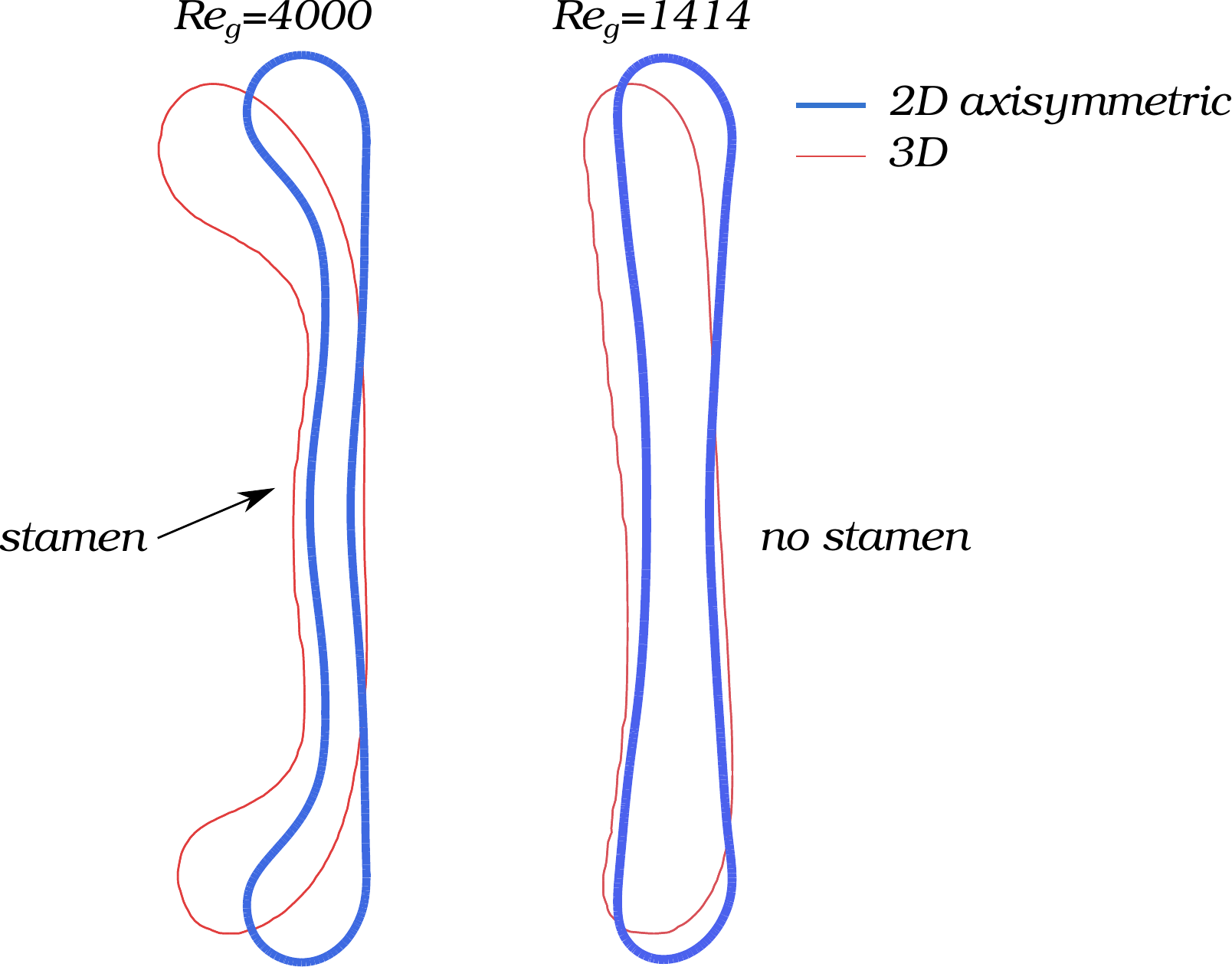}
\centering
\caption{Comparison of drop shapes for $\rho^*=1000$ at $Re_g=1414$ and 4000, $M=1000$, $We=20$ and at $t^*=1.52$ obtained from a 3D and 2D axisymmetric simulation showing a good match between 2D and 3D results.}
\label{fig:flow_shape}
\end{figure}

Although, the flow around the drops in 2D is significantly different from that in 3D, the shapes of the drops are surprisingly the same, and both the 2D and 3D simulations predict the formation of stamen and a forward-bag, for two different Reynolds numbers, as shown in the Figure \ref{fig:flow_shape}. This indicates that the wake structures do not significantly affect the drop morphology and breakup for the drops with higher $\rho^*$ values (lower $\rho^*$ values are investigated in later sections). \citet{Flock2012} also reported a similar conclusion based on the differences in the observations of the PIV realized instantaneous flow fields, showing alternating vortices in the wake region, and the ensemble-averaged of the flow fields, showing a symmetric twin vortex pair in the wake region around the drop.

Another interesting observation from Figures \ref{fig:flow_re4000}-\ref{fig:flow_shape} is that the drop shapes are different for different $Re$ (4000 and 1414) values for the same $M=1000$, $\rho^*=1000$ and $We=20$. Figure \ref{fig:flow_shape} clearly shows the tendency of the drop to form a stamen at $Re=4000$, whereas there is no sign of formation of stamen for $Re=1414$. Further, Table \ref{tab:re_mode} lists the drop breakup modes along with the non-dimensional RT wavenumber, $N_{RT}$, and Table \ref{tab:re_shape} lists the corresponding breakup shapes of the drop for $\rho^*=1000$ at $We=20$, $M=100$ and 1000 and at different $Re$ values. It can be seen that with an increase in $Re$ value for the same viscosity ratio $M$, the drop deformation and hence the breakup mode is changing, effectively altering the breakup transition value of $We$, that is with an increase in $Re$ the breakup transition value of $We$ is decreasing. 

 \begin{table}
 	\begin{center}
  	\begin{tabular}{c|c|c|c|c|c|}
 		\multicolumn{3}{c}{$M=100$} & \multicolumn{3}{c}{$M=1000$} \\ [3pt] 
 		\multirow{4}{*}{$Re$} & 4000 & Bag-Stamen (1.196) & \multirow{4}{*}{$Re$} & 6000 & Bag-Stamen (0.8729) \\ 
 		& 1414 & Bag-Stamen (1.084) &   & 4000 & Bag-Stamen (1.092) \\ 
 		& 500  & Bag (0.71004) &  & 1414 & Bag (0.97) \\ 
 		& 141  & Bag (0.4548) &  &  &  \\ 
 \end{tabular}
 \caption{Breakup modes at different $Re$ values for a drops with $\rho^*=1000$ at $We=20$ and $M=100$ and $1000$. Values in the bracket denotes the $N_{RT}$ values.}
 \label{tab:re_mode}
 \end{center}
 \end{table}
 
 Non-dimensional RT wavenumber, $N_{RT}$, listed in Table \ref{tab:re_mode} also shows an evident increase with increase in $Re$ value, thus reinforcing the argument that the RT-instability is the mechanism of the breakup of drops for high $\rho^*$ values. For different $M$ values, the value of $Re$ for which the the breakup mode changes is also different. Therefore, Ohnesorge number, $Oh$, which varies with $M$ and $Re$ values when everything else is kept constant, is a better parameter to represent this behavior. In terms of Ohnesorge number, a decrease in $Oh$, shifts the transition $We$ to lower values. We note that, although the values of $Oh$ are less than 0.1 (a value below which it is considered that there are no effects of viscous forces on the breakup mechanism), changing the value of $Oh$ still affects the breakup of drops but not to the extent for the values of $Oh>0.1$. This possibly explains the discrepancy in breakup transitional value of $We$ observed by different authors in their experimental results (Table \ref{tab:transitional_we}). 
 

 \begin{table}
 	\begin{center}
 	\begin{tabular}{|c|c|c|c|c|c|}
 		\multicolumn{3}{c}{$M=100$} & \multicolumn{3}{c}{$M=1000$} \\[3pt] \\[-0.5em]
 		\multirow{4}{*}{$Re\ \ \ $} 
 		& 4000 & \raisebox{-0.2in}{ \includegraphics[scale=0.3]{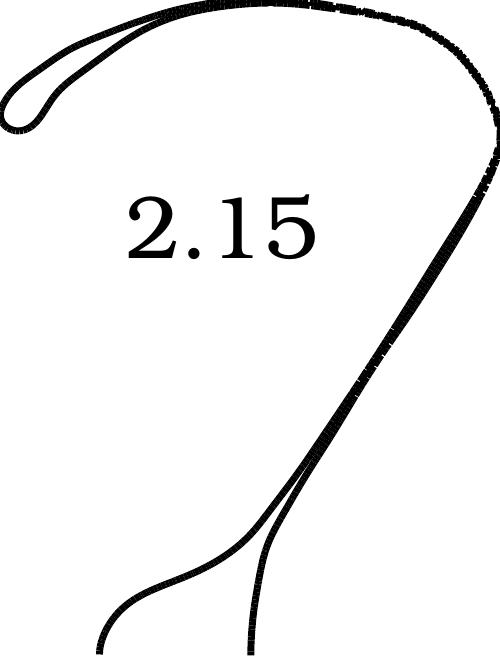}} & \multirow{4}{*}{$Re\ \ \ $} 
		& 6000 & \raisebox{-0.2in}{ \includegraphics[scale=0.4]{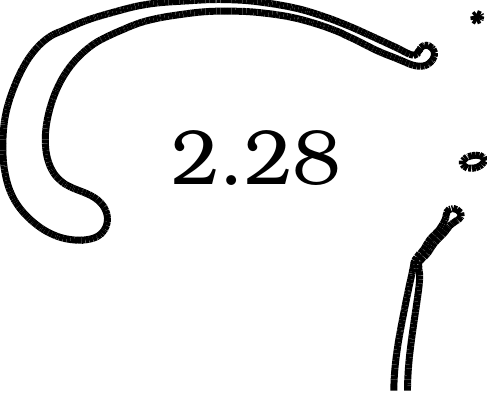}} \\ \cdashline{2-3} \cdashline{5-6} \\[-0.5em]
 		& 1414 & \raisebox{-0.2in}{ \includegraphics[scale=0.3]{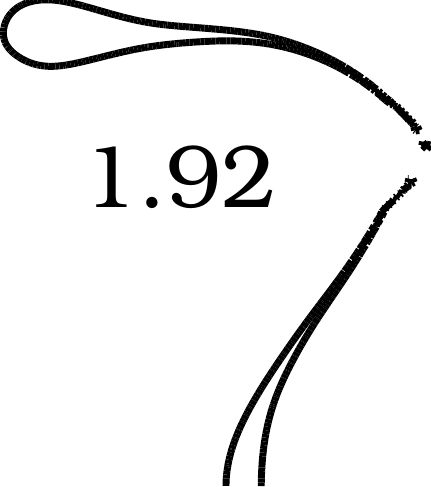} }  &   
 		& 4000 & \raisebox{-0.2in}{ \includegraphics[scale=0.3]{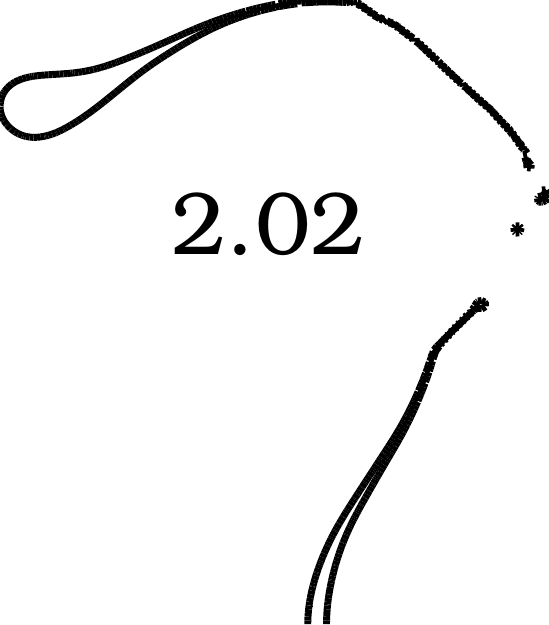} }  \\ \cdashline{2-3} \cdashline{5-6} \\[-0.5em] 
 		& 500  & \raisebox{-0.2in}{ \includegraphics[scale=0.3]{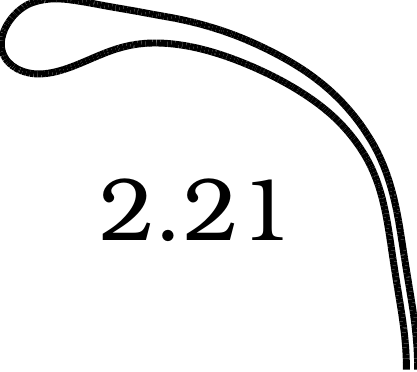} } &  
 		& 1414 & \raisebox{-0.2in}{ \includegraphics[scale=0.3]{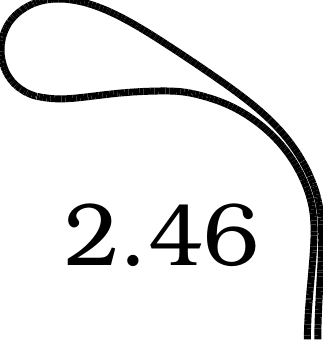} } \\ \cdashline{2-3} \cdashline{5-6} \\[-0.5em] 
 		& 141  & \raisebox{-0.2in}{ \includegraphics[scale=0.3]{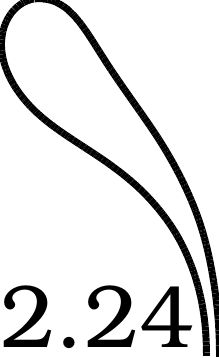} } &  &  &  \\ 
 \end{tabular}
 \caption{Breakup shapes at different $Re$ values for the drops with $\rho^*=1000$ at $We=20$ and $M=100$ and $1000$ along with the time $t^*$ beside them. }
 \label{tab:re_shape}
 \end{center}
 \end{table}

\subsection{Rim dynamics}\label{sec:rim}

For the drops with high $\rho^*$, flow around the drop has relatively low effect on the drop deformation, morphology and breakup and hence the drop deforms into a flat disc and further, RT-instability governs the breakup as already explained in Sections \ref{sec:rt} and \ref{sec:flow}. But for the drops of lower $\rho^*$, the flow field has a greater impact on the drop morphology, deformation and breakup due to higher velocity induced in the drops by the surrounding gas flow as already explained in Section \ref{sec:dcrit} and the flow patterns around the rim guide the direction of alignment of the rim and hence "\textit{rim dynamics}" governs the breakup.   

\begin{figure}
    \centering
    \begin{subfigure}[c]{0.4\textwidth}
        \centering
        \includegraphics[width=\textwidth]{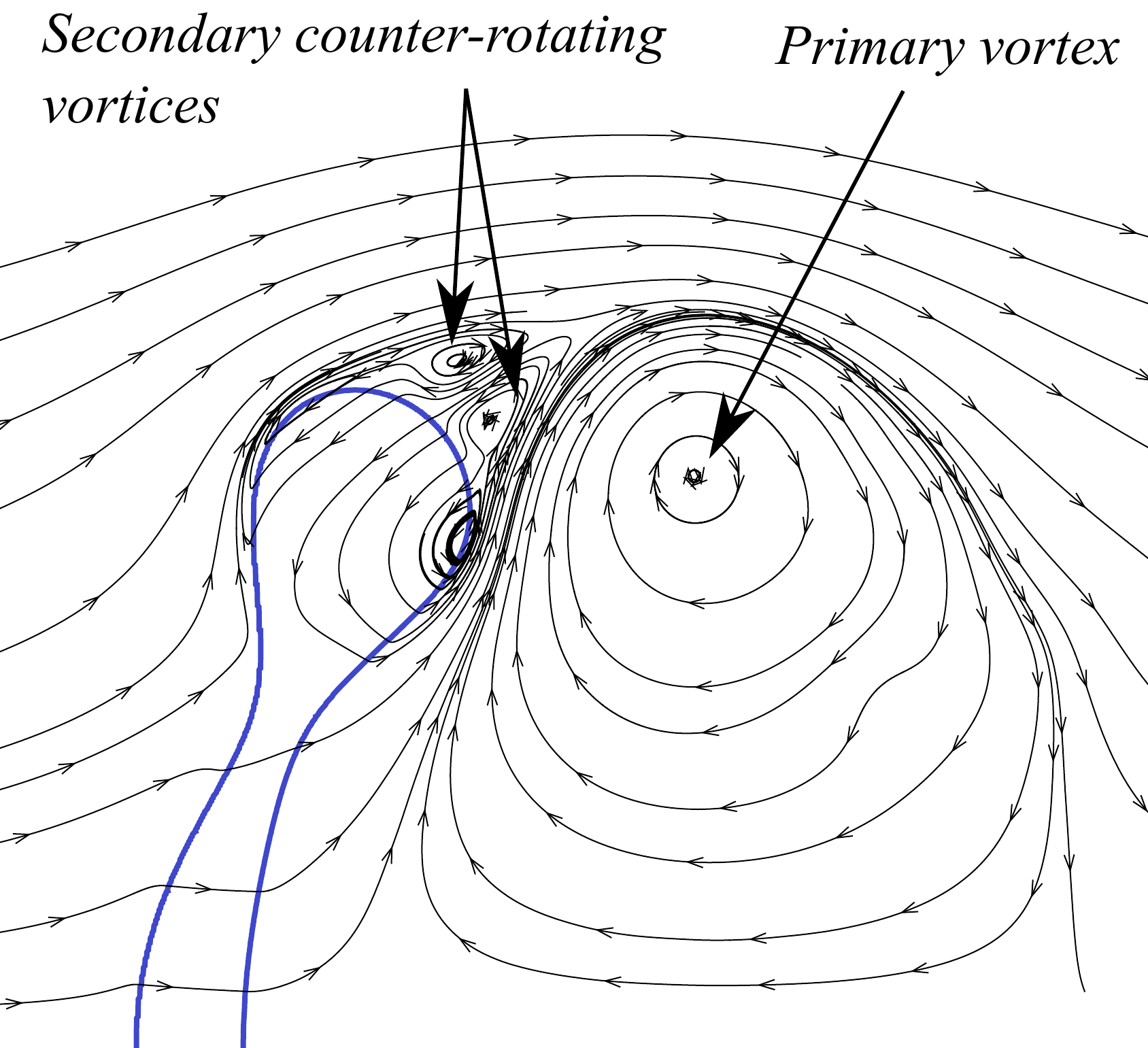}
        \caption{2D axisymmetric simulation. Blue line represents the drop and lines with the arrows are the streamlines of velocity relative to the centroid velocity of the drop.}
        \label{fig:rim_2d_16}
    \end{subfigure}%
    ~ 
    \begin{subfigure}[c]{0.6\textwidth}
        \centering
        \includegraphics[width=0.6\textwidth]{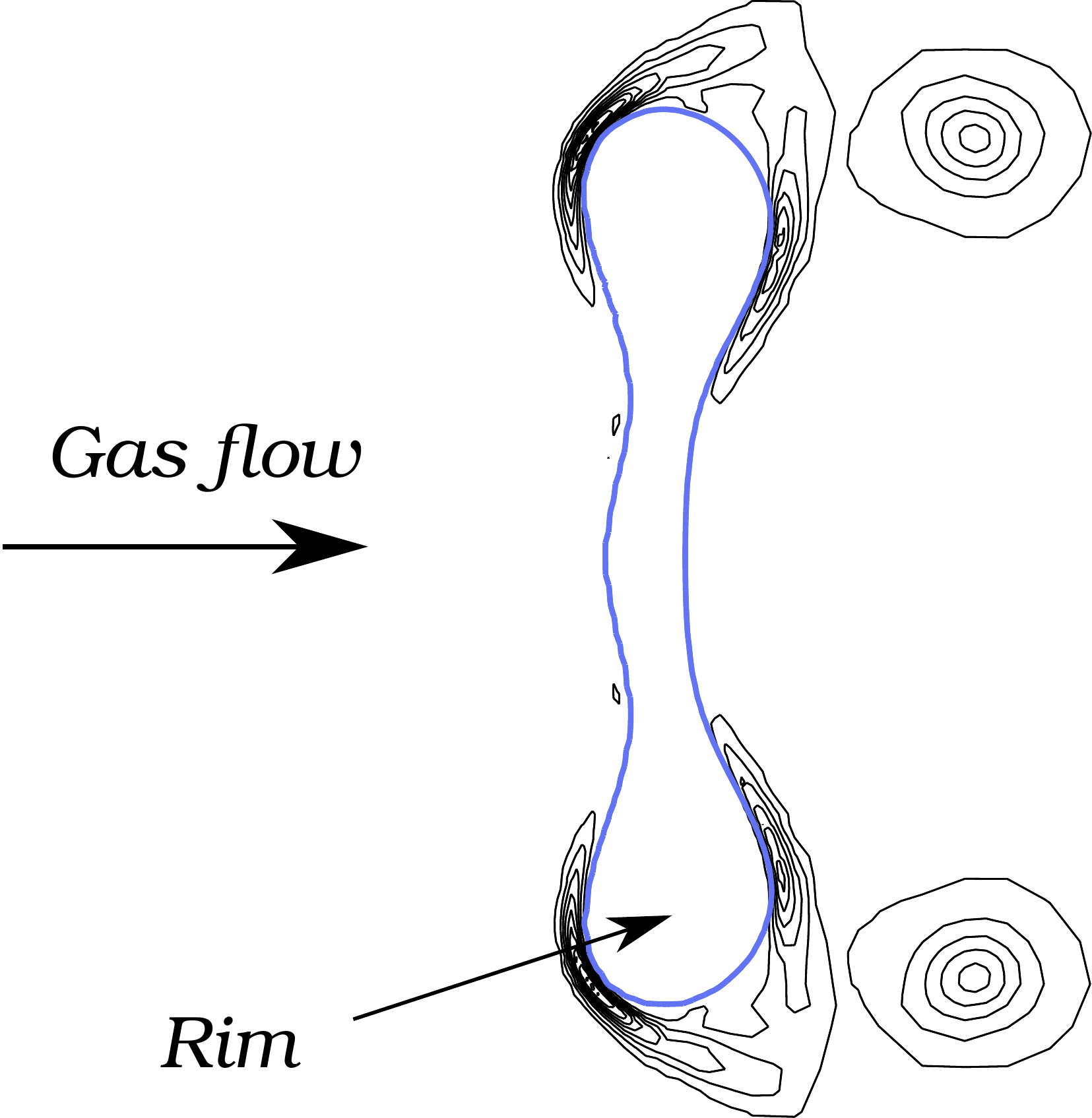}
        \caption{Cross-section view of the 3D simulation. Blue line represents the drop, black lines represents the vorticity contour plot. Arrow represents the direction of gas flow.}
        \label{fig:rim_3d_16}
    \end{subfigure}
    ~
    \begin{subfigure}[c]{0.6\textwidth}
        \centering
        \includegraphics[width=\textwidth]{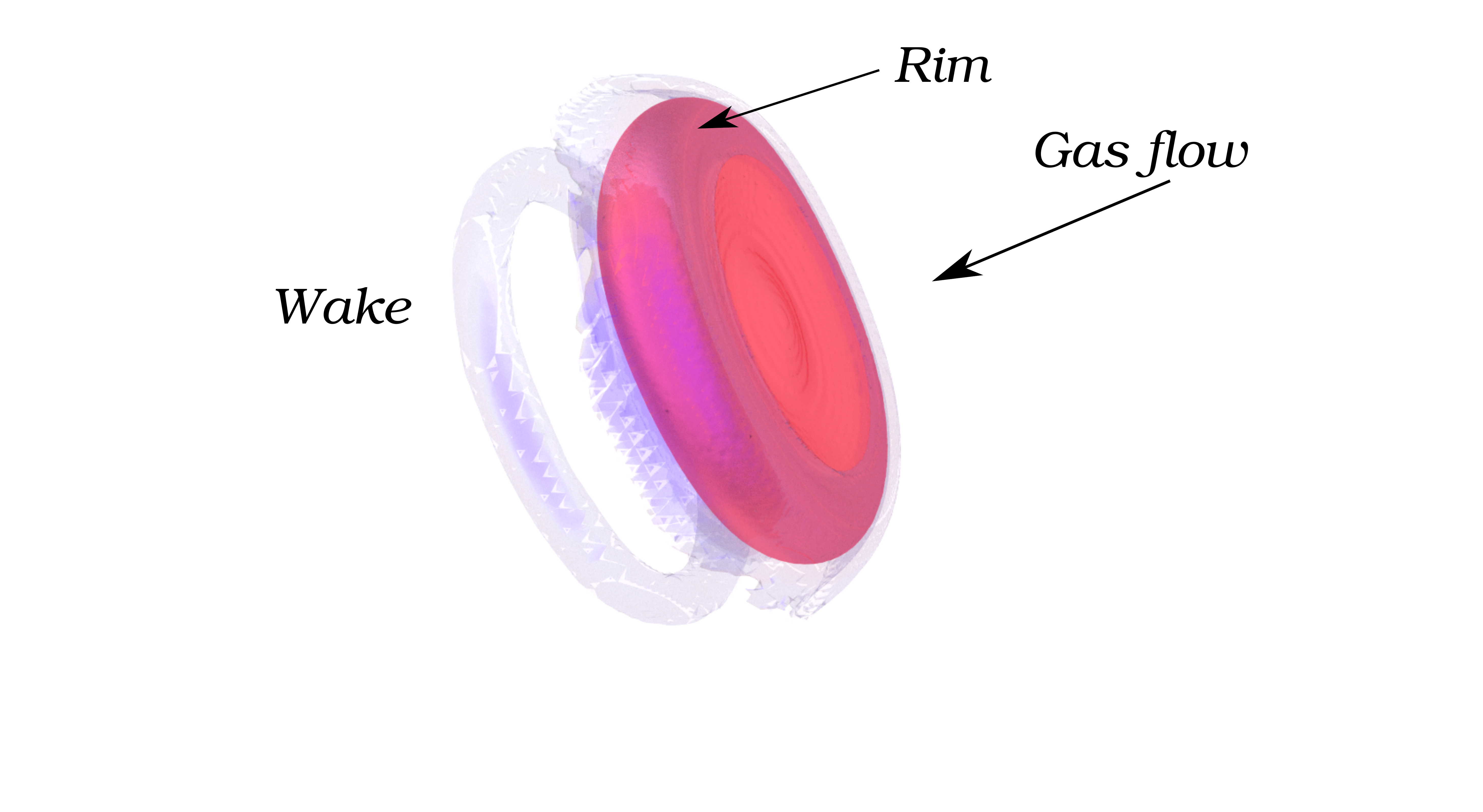}
        \caption{3D simulation in rendered view, showing the drop and the vorticity isosurfaces. Arrow represents the direction of gas flow.}
        \label{fig:render_16}
    \end{subfigure}
    \caption{Comparison of flow field around the drop for $\rho^*=10$ at $Re_g=4000$, $M=100$, $We=20$ and at $t^*=2.02$ obtained from a 3D and 2D axisymmetric simulation.}
    \label{fig:rim_16}
\end{figure}

\begin{figure}
    \centering
    \begin{subfigure}[c]{0.4\textwidth}
        \centering
        \includegraphics[width=\textwidth]{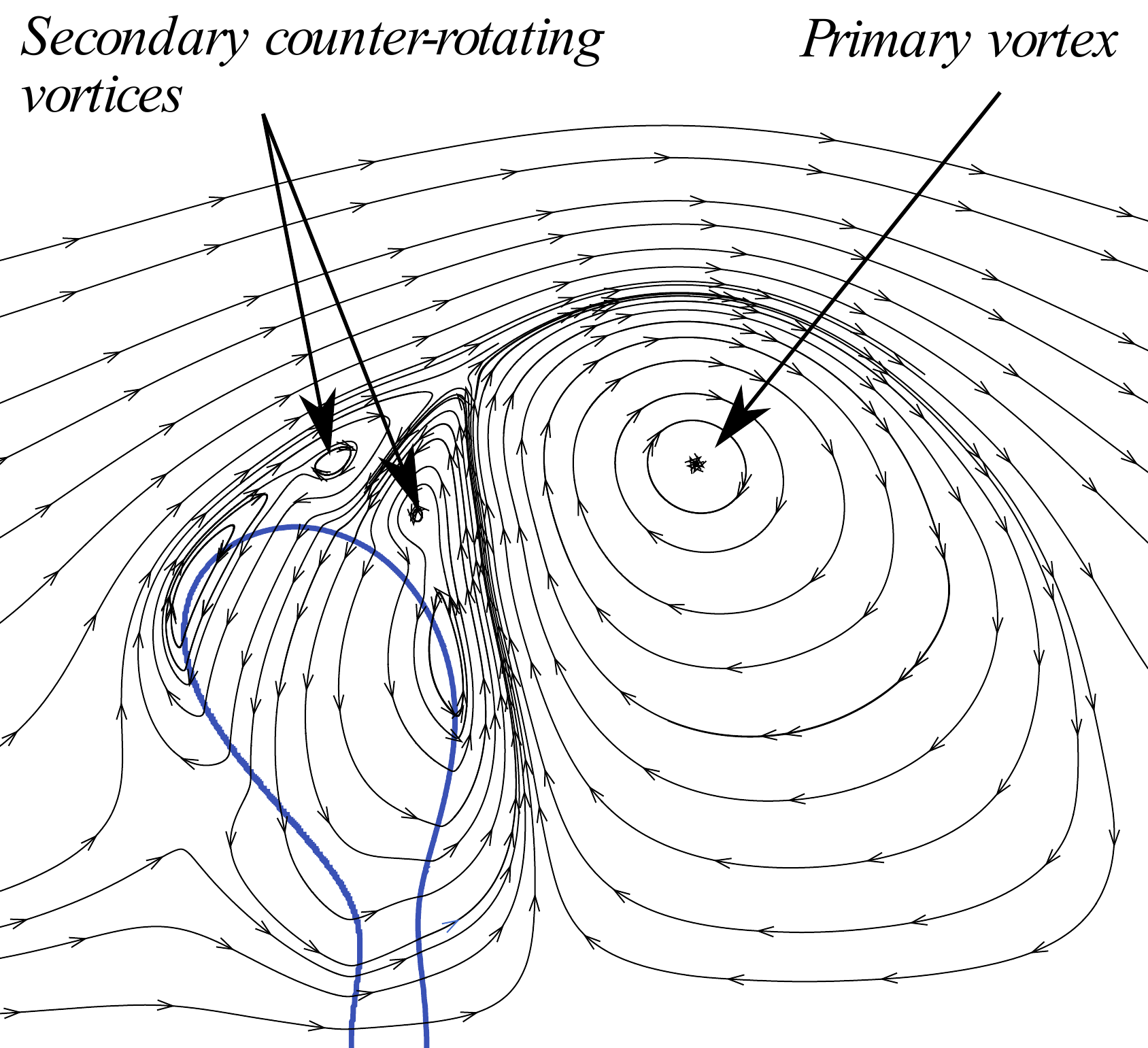}
        \caption{2D axisymmetric simulation. Blue line represents the drop and lines with the arrows are the streamlines of velocity relative to the centroid velocity of the drop.}
        \label{fig:rim_2d_20}
    \end{subfigure}%
    ~ 
    \begin{subfigure}[c]{0.6\textwidth}
        \centering
        \includegraphics[width=0.6\textwidth]{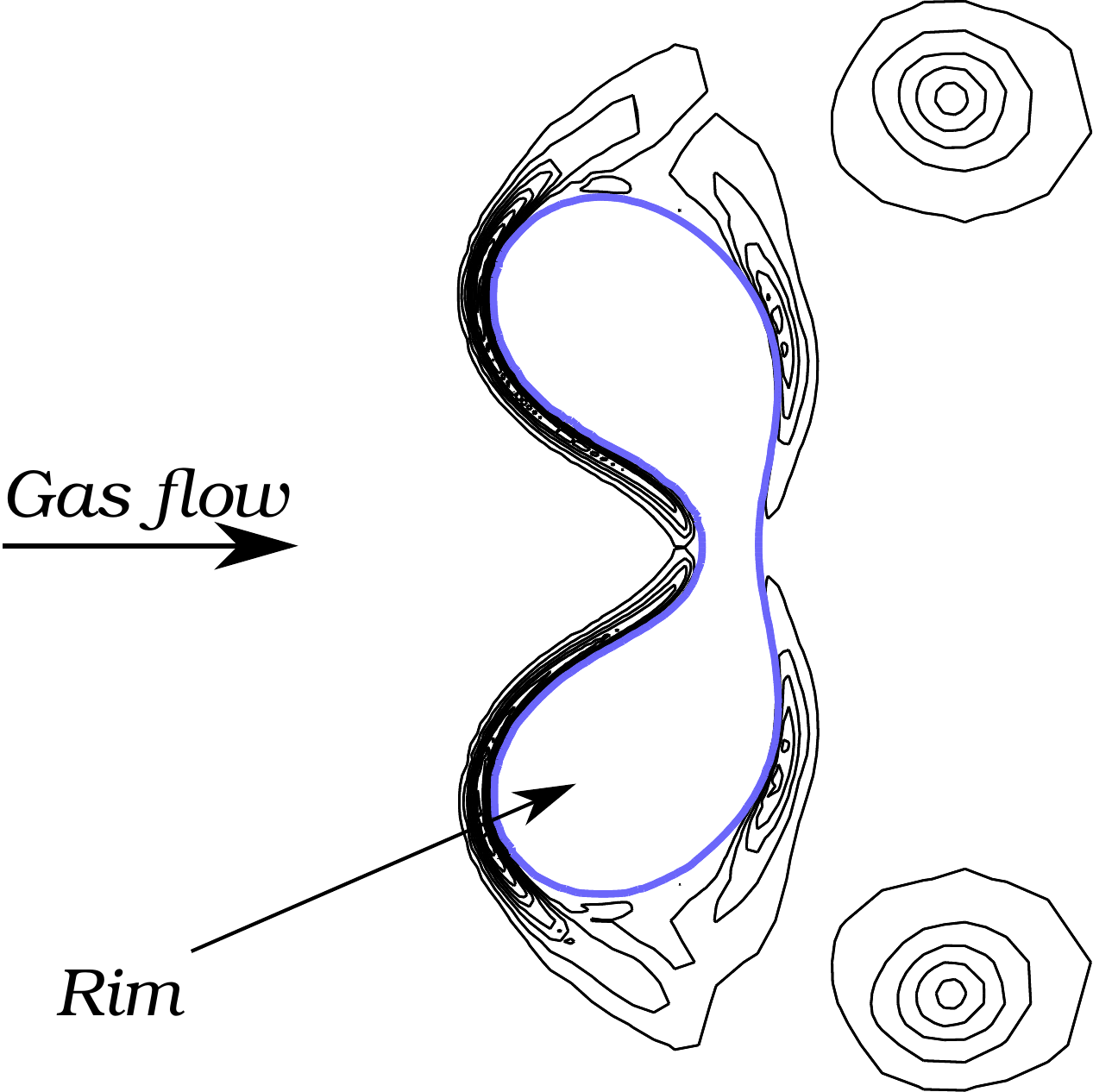}
        \caption{Cross-section view of the 3D simulation. Blue line represents the drop, black lines represents the vorticity contour plot. Arrow represents the direction of gas flow.}
        \label{fig:rim_3d_20}
    \end{subfigure}
    ~
    \begin{subfigure}[c]{0.6\textwidth}
        \centering
        \includegraphics[width=\textwidth]{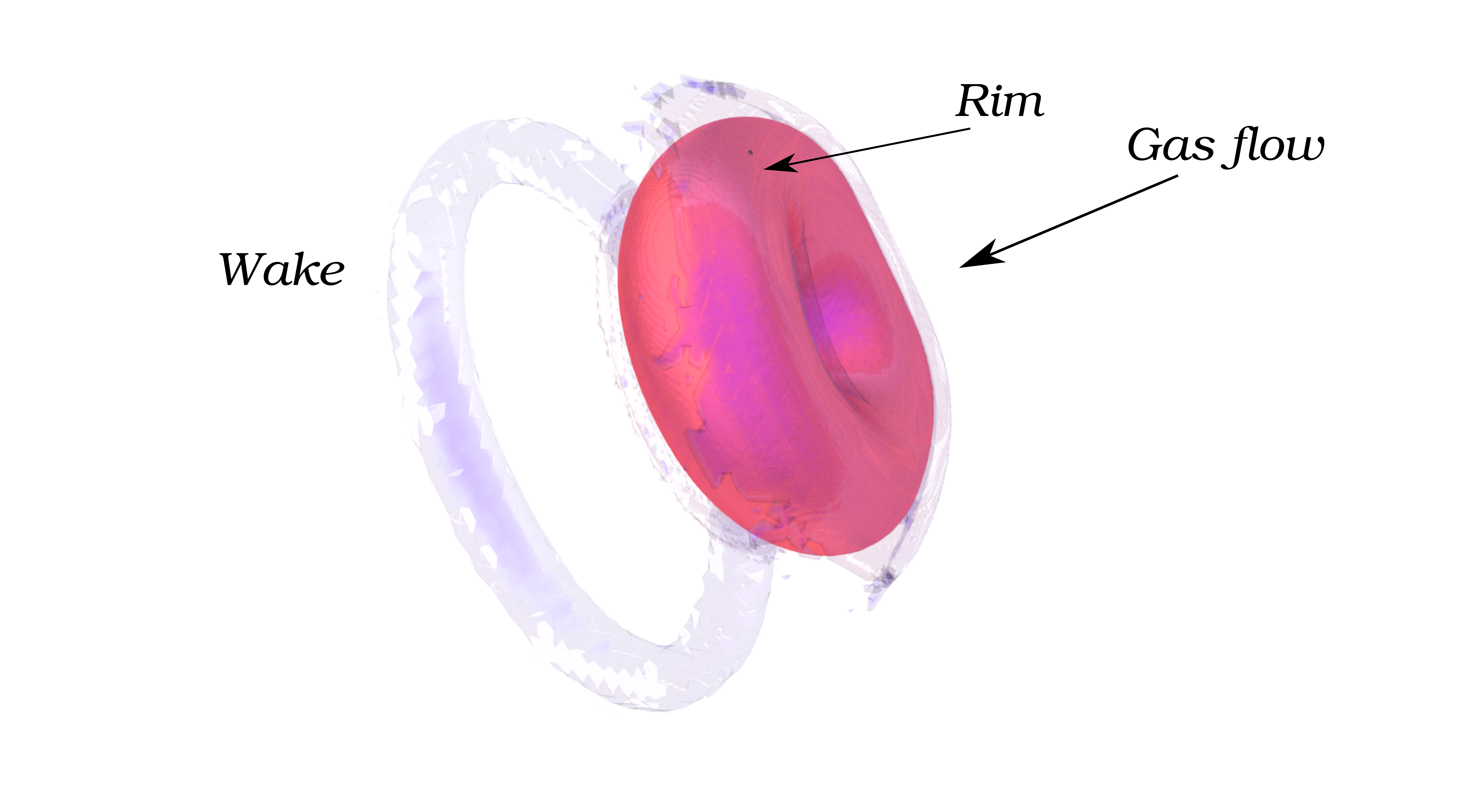}
        \caption{3D simulation in rendered view, showing the drop and the vorticity isosurfaces. Arrow represents the direction of gas flow.}
        \label{fig:render_20}
    \end{subfigure}
    \caption{Comparison of flow field around the drop for $\rho^*=10$ at $Re_g=4000$, $M=100$, $We=20$ and at $t^*=2.53$ obtained from a 3D and 2D axisymmetric simulation.}
    \label{fig:rim_20}
\end{figure}

Figures \ref{fig:rim_16} and \ref{fig:rim_20} show the 2D axisymmetric and 3D simulations of the drop for $\rho^*=10$, $M=100$ at $We=10$, $Re_g=4000$, $t^*=2.02$ and at $t^*=2.53$. Interestingly, unlike the drops with high $\rho^*$, a turbulent vortex shedding is not seen in the wake of the drop in 3D simulations. Instead undisturbed vortex rings are seen in both 3D and 2D axisymmetric simulations. Since a 2D axisymmetric simulation predicts drop shape as well as the flow around the drop accurately, the initial phase of the deformation of the drop during secondary breakup for low $\rho^*$ values can indeed be considered axisymmetric.

The primary vortex ring, seen in Figures \ref{fig:rim_2d_16} and \ref{fig:rim_2d_20}, is formed on the leeward surface of the drop due to the separation of an initially attached boundary layer on the surface of the drop. This vortex ring pulls the rim of the drop along resulting in the formation of a backward-bag. Eventually this vortex ring  pinches-off from the boundary layer and moves downstream, with respect to the drop, along with the formation of a pair of secondary counter-rotating vortex rings at the rim (Figures \ref{fig:rim_2d_16} and \ref{fig:rim_2d_20}) due to the flow separation behind the rim. This counter-rotating vortex rings induce opposite directional velocity in the rim of the drop and deflects it more towards the upstream direction turning the drop into a forward-bag. 

At later times, this primary vortex ring becomes strongly asymmetrical and eventually sheds vortex at the vortex formation time of $t/t_{cv}=4$ (as discussed earlier in Section \ref{sec:flow}). For example, $(U_g-u_{drop})/u_g=0.45$ for the case shown in Figure \ref{fig:rim_16} and hence the asymmetric vortex formation is expected to occur at $t^*\sim3.04$. But according to our observations, based on our simulations, vortex ring never reaches an asymmetrical state because the relative velocity between the drop and the gas, $U_g-u_{drop}$, which feeds into the circulation of the vortex ring reduces substantially with time, and hence the critical circulation required for asymmetry in the vortex rings is never achieved for the drops with low $\rho^*$ values. In contrast, for high density ratios $U_g-u_{drop}$ reduces relatively slowly with time and hence the critical circulation is quickly achieved at very early stages of the deformation of the drop that leads to asymmetric vortex rings and eventually a turbulent wake is observed.

\section{Summary and Conclusions} \label{sec:conclude}

In the present study, we performed fully resolved numerical simulations of a drop in a high speed gas flow to investigate the effect of density ratio and Reynolds number on the secondary breakup of the drops. These simulations were performed for a moderate Weber number range (20-120), where bag breakup, multi-mode and sheet-thinning breakup modes have been observed in experiments. 

Previous studies reported conflicting views on the effect of density ratio on the breakup modes and drop morphology \citep{Aalburg2003,Kekesi2014,Yang2016}. To resolve these discrepancies, we performed a large set of simulations with different values of $\rho^*$ from 10 to 1000, and $We$ from 20 to 120. Further, we vary $Re$ and $M$ independently to delineate their effects on drop morphology. 

In what follows, we present the important conclusions from this study.
\begin{enumerate}
\item For high $\rho^*$ values, drops deform into a flat disc, whereas for low $\rho^*$ values the drops do not deform into a flat disc at all, instead they bend towards the downstream direction and for intermediate $\rho^*$ values, there is a gradual variation in the bend.

\item Axial and radial components of the velocities at the center of the drop, $u_{center}$, and at the rim, $u_{rim}$, decrease with an increase in $\rho^*$ values, which also follows from the scaling relation based on momentum transfer $U_l\sim\sqrt{1/\rho^*}U_g$. Further, the difference in axial velocity, ($u_{center}-u_{rim}$), decreases with an increase in the $\rho^*$ values, which explains the higher stretching and bending of the drops for low $\rho^*$ values. 

\item Displacement of the drops decreases with increase in $\rho^*$ values. This is due to the differences in the centroid velocity of the drop, $u_{drop}$, and the momentum transferred to the leeward side of the drops, which depends on the kinematic viscosity of the drop, $\nu_l$.

\item Breakup time, $t^*_b$, and the distance travelled in the streamwise direction, $x_l/d_0$, at the onset of breakup are higher for the drops with $\rho^*=10$ than for the drops with $\rho^*=50-1000$ and they decrease with an increase in $We$, whereas relative velocity, $u_r=(u_g-u_l)/u_g$ has a continuous variation from $\rho^*=50-1000$ and for high density ratios the values are in good agreement with the experimental observations of \citet{Dai2001,Zhao2010}.

\item Drops for $\rho^*=10$ at $We=20$ and $40$ and for $\rho^*=50$ at $We=20$ do not breakup at all, which is also in agreement with the simulations of \citet{Han2001}. We explain this using the instantaneous Weber number, $We_{inst}=\rho_g(U_g-u_{drop})^2d_0/\sigma$. The values of $We_{inst}$ for these conditions are less than $We_{crit}$ at the onset of breakup or the time where there could have been a pinch-off at the thinnest section of the drop, whereas $We_{inst}>We_{crit}$ at the onset of breakup for all other cases where breakup is observed. 

\item Interesting drop shapes are observed with varying $\rho^*$ values keeping $We$ constant. At $We=20$, a forward-bag is seen at $\rho^*=10$, transient  canopy-top shape for $\rho^*=50$ and 100 and forward-bag with stamen for $\rho^*\ge150$. The size of the stamen decreases with increase in $\rho^*$ value. At $We=40$ and higher, backward-bag is seen for $\rho^*=10$, whiplash with sheet-thinning for $\rho^*=50$ and sheet-thinning for $\rho^*\ge100$. The rim is thicker for low $\rho^*$ values, which could be due to higher Taylor-Culick velocity for low $\rho^*$ values.

\item In addition to the differences in deformation, breakup morphology  and breakup modes, the breakup mechanism is also different for higher and lower $\rho^*$ values. At higher $\rho^*$ values, the formation of bag is due to the Rayleigh-Taylor instability. The non-dimensional RT wavenumber, $N_{RT}$ in the cross-stream direction of the drop agrees very well with the range proposed by \citet{Zhao2010}. Numerical and theoretically calculated non-dimensional growth rates, $\omega^*$, on the other hand are not in direct quantitative agreement with the theoretical estimated for the instability on a planar surface; but they are proportional to the theoretical growth rate, implying that there could be "\textit{end-effects}" associated with the growth of RT wave.

\item Study of the flow around the drop, in 3D simulations, reveals that the vortex ring formed due to flow separation in the wake region of the drop, develops asymmetries and sheds leading to the formation of turbulent wake region and the time taken to form these asymmetries agrees very well with the vortex formation time scales. However, vortex ring observed in 2D axisymmetric simulations is stable and never develops any asymmetry. Nevertheless, the drop shapes are same in 2D and 3D cases, implying that the flow has only a weak effect on the drop shape for higher $\rho^*$ values, which is in agreement with the experimental observations of \citet{Flock2012}.

\item Increasing gas Reynolds number, $Re$ alters the breakup and we see that the drop breakup mode transitions from a bag to bag-with-stamen for an increase of $Re$ from 1414 to 4000 at viscosity ratio $M=1000$. Non-dimensional RT wavenumber, $N_{RT}$, also increases with an increase in $Re$, and the $N_{RT}$ values again conform to the range given by \citet{Zhao2010}, indicating that the RT-instability is indeed the breakup mechanism at higher $\rho^*$ values. The effect of $M$ is that it shifts the value of transition value of $Re$ (critical $Re$ across which there is a change in the breakup mode for a given $We$). Hence, combining the effect of $M$ and $Re$, Ohnesorge number, $Oh$, is a better parameter to represent this behavior. This explains the discrepancy in breakup transitional values of $We$ observed by different authors in their experiments (Table \ref{tab:transitional_we}).

\item For lower $\rho^*$ values, breakup is governed by the dynamics of the rim. Flow around the drop has a greater impact (in comparison to high $\rho^*$ drops) on the drop shape, deformation and breakup due to higher velocity induced in the drop by the surrounding gas flow. For low $\rho^*$, unlike for higher $\rho^*$ values,  the vortex ring formed at the rim never develops asymmetries until the breakup of the drop.

\end{enumerate}

To conclude, the drops for $\rho^*<150$ behave differently from $\rho^*\ge150$ at the same $We$, making "\textit{Density ratio}" an important parameter in characterizing secondary breakup of drops and also in the study of liquid jets in gas crossflow.  The present study, describes the differences in the behavior of the drops for different $\rho^*$ values, such as in the air-water system in atmospheric conditions, where the $\rho^*\sim1000$, and in high pressure applications where $\rho^*$ can be $\sim100$, and also in the manufacturing of pellets by quenching molten metal in a pool of cold water where $\rho^*\sim 1-10$.
    
In the present work, we have essentially focused on the drop deformation and breakup modes in the moderate Weber number regime. Based on our results, we believe that a similar systematic study on vibrational mode of breakup as well as high Weber number breakup would reveal interesting effects of the density ratio on the deformation and breakup mechanisms.



\subsection*{Acknowledgements} The authors acknowledge the guidance from Prof. B.N Raghunandan (Department of Aerospace Engineering, Indian Institute of Science, Bangalore) in conducting the research work presented in this article.

\subsection*{Declarations of interest} None.


\subsection*{Funding} No direct funding was used to perform this work.

\appendix

\section*{Appendix A: Relation between non-dimensional growth rate and instantaneous Bond number} \label{app:growthrate}

Non dimensional growth rate is given by,

\begin{equation}
    \omega^* = \omega\frac{d\sqrt{\rho^*}}{U_g}
\end{equation}

From equation \ref{eq:growthrate}, 

\begin{equation}
    \omega^* = \sqrt{ka\Big(\frac{\rho^*-1}{\rho^*+1}\Big)}\frac{d\sqrt{\rho^*}}{U_g}
\end{equation}

Substituting for wavenumber, $k=\sqrt{{\rho_la}/{3\sigma}}$ and rearranging, 

\begin{equation}
    \omega^* = \frac{1}{3^{\frac{1}{4}}}\frac{{At}^{\frac{1}{2}}{Bo}^{\frac{3}{4}}}{\sqrt{We}}
\end{equation}

Here $At$ is the Atwood number. Hence the $\omega^*$ scales as $Bo^\frac{3}{4}$ for a given $\rho^*$ and $We$.

 \bibliographystyle{elsarticle-harv}


\end{document}